\documentclass{aa}  
\usepackage{graphicx}
\usepackage[T1]{fontenc}
\usepackage{txfonts}
\usepackage{epstopdf}
\usepackage{newtxtext,newtxmath}
\newcommand{\asec}{$^{\prime\prime}$}

\def\SigmaH2{$\Sigma $(${\rm H_2}$)}
\def\r1415{$^{14}$N/$^{15}$N}

\def\cyclic{{\it c-}C$_3$H$_2$}

\def\H{N$_{2}$H$^{+}$}

\def\15N{$^{15}$NNH$^+$}
\def\N15{N$^{15}$NH$^+$}

\def\FORM{H$_2$CO}

\def\HCOp{\mbox{HCO$^+$}}
\def\HCOpI{\mbox{H$^{13}$CO$^+$}}

\def\METH{CH$_3$OH}

\def\H13CN{\mbox{H$^{13}$CN}}

\def\kms{\mbox{km~s$^{-1}$}}
\def\cmc{cm$^{-3}$}

\def\kms{km\,s$^{-1}$}

\begin{document} 

   \title{CHEMOUT: CHEMical complexity in star-forming regions of the OUTer Galaxy. I. Organic molecules and tracers of star-formation activity}
   
   \author{F. Fontani
          \inst{1,2} 
          \and
          L. Colzi\inst{3,1}
          \and
          L. Bizzocchi\inst{2,4}
          \and
          V.M. Rivilla\inst{3,1}
          \and
          D. Elia\inst{5}          
          \and
          M.T. Beltr\'an\inst{1} 
          \and
          P. Caselli\inst{2}
          \and
          L. Magrini\inst{1}
          \and
          A. S\'anchez-Monge\inst{6}
          \and 
          L. Testi\inst{7,1}
          \and
          D. Romano\inst{8}
          }

   \institute{INAF-Osservatorio Astrofisico di Arcetri, Largo E. Fermi 5, I-50125, Florence, Italy\\
            \email{francesco.fontani@inaf.it}
            \and
            Centre for Astrochemical Studies, Max-Planck-Institute for Extraterrestrial Physics, Giessenbachstrasse 1, 85748 Garching, Germany
            \and
            Centro de Astrobiolog\'ia (CSIC-INTA), Ctra. de Ajalvir Km. 4, Torrej\'on de Ardoz, 28850 Madrid, Spain
            \and
            Dipartimento di Chimica ''Giacomo Ciamician'', Universit\`a di Bologna, Bologna, Italy
            \and
            INAF - IAPS, via Fosso del Cavaliere, 100, I-00133 Roma, Italy
            \and
            I. Physikalisches Institut, Universit\"at zu K\"oln, Z\"ulpicher Str. 77, 50937 K\"oln, Germany
            \and
            European Southern Observatory, Karl-Schwarzschild-Str. 2, 85748 Garching, Germany
            \and
            INAF, Osservatorio di Astrofisica e Scienza dello Spazio, Via Gobetti 93/3, 40129, Bologna, Italy
             }

 \date{Received ; accepted }

 
 \abstract
   {The outer Galaxy is an environment with metallicity lower than the Solar one. Because of this,
   the formation and survival of molecules in star-forming regions located in the inner and outer Galaxy
   is expected to be different.}
   {To gain understanding on how chemistry changes throughout the Milky Way, it is crucial to observe outer 
   Galaxy star-forming regions to constrain models adapted for lower metallicity environments.}
   {In this paper we present a new observational project: chemical complexity in star-forming regions
   of the outer Galaxy (CHEMOUT). The goal is to unveil the chemical composition in 35 dense molecular
   clouds associated with star-forming regions of the outer Galaxy through observations obtained with 
   the IRAM 30m telescope in specific 3mm and 2mm spectral windows.}
   {In this first paper, we present the sample, and report the detection at 3~mm of simple organic
   species \HCOp, \HCOpI, HCN, \cyclic, HCO, C$_4$H, and HCS$^+$, of the complex hydrocarbon 
   CH$_3$CCH, and of SiO, CCS and SO. 
   From the optically thin line $J_{K_a,K_b}=2_{1,2}-1_{0,1}$ of  \cyclic\  we estimate new 
   kinematic heliocentric and Galactocentric distances based on an updated rotation curve of the Galaxy.
   The detection of the molecular tracers does not seem to have a clear dependence on the Galactocentric
   distance. Moreover, with the purpose of investigating the occurrence of outflows and investigate the association 
   with protostellar activity, we analyse the \HCOp\ line profiles. We find high velocity 
   wings in $\sim 71\%$ of the targets, and their occurrence does not depend on the Galactocentric
   distance.}
   {Our results, confirmed by a statistical analysis, show that the presence of organic molecules and tracers
   of protostellar activity is ubiquitous in the low-metallicity environment of the outer Galaxy. Based on this, 
   and on the additional evidence that small, terrestrial planets are omnipresent in the Galaxy, we support
   previous claims that the definition of Galactic Habitable Zone should be rediscussed in view of the 
   ubiquitous capacity of the interstellar medium to form organic molecules.}


\titlerunning{CHEMOUT: chemical complexity in the outer Galaxy. I}

\maketitle
%

\section{Introduction}
\label{intro}

\begin{table*}
\begin{center}
\caption{Source parameters extracted from Table~1 of Blair et al.~(2008).}
\begin{tabular}{lccccccc}
\hline
source &  R.A. & Dec. & $V_{\rm LSR}$$^{(1)}$ &  $N(H_2)$$^{(2)}$ & $R^{old}_{\rm GC}$$^{(3)}$ & $M$$^{(4)}$ & $L$$^{(4)}$ \\
            & (J2000) & (J2000)  & \kms\ & $\times 10^{21}$cm$^{-2}$ & kpc & M$_{\odot}$ & ($\times 10^3$) L$_{\odot}$ \\
\hline
WB89-315  & 00:05:53.8 & 64:05:17    & --95.09 & 3.7 & 17.1     & & \\
WB89-379 & 01:06:59.9 & 65:20:51     & --89.32 & 6.5 & 17.3     & 532 & 1.44 \\
WB89-380 & 01:07:50.9 & 65:21:22	   & --86.67 & 11.4 & 17.0   & &  \\
WB89-391 & 01:19:27.1 & 65:45:44	   & --86.06 & 5.2 & 16.9     & &   \\
WB89-399 & 01:45:39.4 & 64:16:00      & --82.19 &  6.3 & 16.8   & 1685 & 16.3 \\
WB89-437 & 02:43:29.0 & 62:57:08     & --71.72 & 14.2 & 16.2   & &   \\
WB89-440 & 02:46:07.3 & 62:46:31     & --72.20 & 4.1  & 16.4    & & \\
WB89-501 & 03:52:27.6 & 57:48:34     & --58.44 & 11.2 & 16.4   & &  \\
WB89-529 & 04:06:25.5 & 53:21:49       & --60.08 & 4.7  & 17.8  & 408 & 10.1 \\
WB89-572 & 04:35:58.3  & 47:42:58     & --48.03 & 3.8  & 20.4   & & \\
WB89-621 & 05:17:13.4  & 39:22:15     & --25.38 & 13.0 & 22.6  & 920 & 41.2 \\
WB89-640 & 05:25:40.7 & 41:41:53     & --25.42 & 3.2 & 18.4     & &  \\
WB89-670 & 05:37:41.9 & 36:07:22       & --17.52 & 7.3 & 23.5   & &  \\
WB89-705 & 05:47:47.6 & 35:22:01      & --12.10 & 1.7 & 21.4    & &  \\
WB89-789 & 06:17:24.3 & 14:54:37     & 34.33 & 5.8 & 20.3       & 1066 & 9.75 \\
WB89-793 & 06:18:41.7 & 15:04:52     & 30.48 & 5.9 & 18.1       & 204 & 0.77 \\
WB89-898 & 06:50:37.3 & $-$05:21:01    & 63.41 & 2.5 & 16.4   & &  \\
19383+2711 & 19:40:22.1 & 27:18:33     & --66.85 & -- & 13.2     & &  \\
19423+2541 & 19:44:23.2 & 25:48:40    & --72.62 &  -- & 13.6     & 1278 & 104.5  \\
19489+3030 & 19:50:53.2 & 30:38:09      & --68.91 & -- & 13.0    & & \\
19571+3113 & 19:59:08.5  & 31:21:47     & --62.48 & -- & 12.2    & & \\
20243+3853 & 20:26:10.8  & 39:03:30     & --73.06 & --  & 12.9   & 891 & 15.9 \\
WB89-002 & 20:37:22.3 & 47:13:54	        & --2.75 & 7.8 &  8.6   & & \\
WB89-006 & 20:42:58.2 & 47:35:35	       & --91.38 & 6.3 & 14.9  & &  \\
WB89-014 & 20:52:07.8 & 49:51:28         & --96.03 & 4.6  & 15.5 & & \\
WB89-031 & 21:04:18.0 & 46:53:10         & --89.40 & 1.2  & 14.6 & & \\
WB89-035 & 21:05:19.7 & 49:15:59	       & --77.68 & 5.2 & 13.4  & 367 & 5.23 \\
WB89-040 & 21:06:50.0 & 50:02:09           & --62.65 & 4.1 & 12.1 & 613 & 0.62 \\
WB89-060 & 21:15:56.0 & 54:43:33	         & --84.29 & 9.3 & 14.0 & &  \\
WB89-076 & 21:24:29.0 & 53:45:35	         & --97.17 & 5.0 & 15.7 & 355 & 0.82 \\
WB89-080 & 21:26:29.1 & 53:44:11	         & --74.24 & 8.5 & 13.1 & 299 & 1.44 \\
WB89-083 & 21:27:47.7 & 54:26:58	         & --93.77 & 2.8 & 15.3 & 220 & 0.82 \\
WB89-152 & 22:05:15.4 & 60:48:41         & --88.13 & 2.8  & 14.8 & & \\
WB89-283 & 23:32:23.8 & 63:33:18	        & --94.49 & 5.8 & 16.5 & 140 & 4.80 \\
WB89-288 & 23:36:08.1 & 62:23:48        & --100.87 & 3.4 & 17.5 & 373 & 3.22 \\
\hline
\end{tabular}
\end{center}
$^{(1)}$ Local Standard of Rest velocities used to centre the spectra; \\
$^{(2)}$ H$_2$ column densities from CO (1--0) (Blair et al.~\citeyear{blair08}), assuming a standard CO-H$_2$
conversion factor of~1.8$\times 10^{20}$ cm$^{-2}$ (K Km s$^{-1}$)$^{-1}$ (Dame et al.~\citeyear{dame01})
valid at the Solar Circle. The values are averaged within the Arizona Radio Observatory (ARO) main beam of 44\asec; \\
$^{(3)}$ Galactocentric distances based on the rotation curve of Brand \& Blitz~(\citeyear{beb93}). 
We will derive updated $R_{\rm GC}$ in Sect.~\ref{dist}; \\
$^{(4)}$ total gas mass, $M$, and bolometric luminosity, $L$, respectively, derived from {\em Herschel} measurements 
by Elia et al.~(\citeyear{elia21}), and rescaled when needed to the new heliocentric distances calculated in this
work (Table~\ref{tab:newdist}). \\
\label{tab:sources}
\end{table*}

The outer Galaxy (OG) is the portion of the Galactic disk located at Galactocentric distances, $R_{\rm GC}$, 
in between $9\leq R_{\rm GC}\leq 24$~kpc, i.e. beyond the Solar Circle. 
It shows chemical properties significantly different from those of the inner Galaxy. In particular, the overall 
metallicity is lower than the Solar one (e.g., a factor of four lower at $R_{\rm GC}=19$~kpc, 
Shimonishi et al.~\citeyear{shimonishi21}). The elemental abundances of oxygen, carbon, and nitrogen, i.e. the three most abundant 
elements in the Universe after hydrogen and helium, decrease as a function of $R_{\rm GC}$ (see e.g. Esteban et 
al.~\citeyear{esteban17}), as all the other elements, following the so-called radial metallicity gradient. Despite the fact
that the OG is believed to be more favourable than the inner Galaxy for preserving life on habitable planets, due 
to the low rate of disruptive events (e.g. Piran \& Jim\'enez et al.~\citeyear{pej14}; Vukoti\'c et al.~\citeyear{vukotic16}),
the lower abundance of heavy elements with respect to the Solar one has 
suggested in the past that this zone is not suitable to form planetary systems in which Earth-like planets could be born and 
might be capable of sustaining life (Prantzos~\citeyear{prantzos08}, Ram\'irez et al.~\citeyear{ramirez10}). Chemical 
evolution models predict that the so-called Galactic Habitable Zone (GHZ) in the Milky Way is an annulus extended up to 
$R_{\rm GC}\sim 9$~kpc, with maximum habitability at $R_{\rm GC}\sim 8$~kpc (Spitoni et al.~\citeyear{spitoni14}, \citeyear{spitoni17}). 

However, this scenario has been challenged by recent observational results, in which the occurrence of Earth-like planets 
does not seem to depend on the Galactocentric distance (e.g.~Maliuk \& Budaj~\citeyear{meb20}).
The formation of small, terrestrial planets does not require a metal-rich environment, suggesting that their existence might be 
widespread in the disk of the Galaxy (Mulders~\citeyear{mulders18}). In addition, the mass of  Super-Earths and Sub-Neptunes 
planets is not determined by the availability of solids, but is instead regulated by poorly known processes, with a very weak 
dependence on metallicity (Kutra et al.~\citeyear{kutra21}, Pacetti et al.~\citeyear{pacetti20}).

All this indicates that, even at metallicities lower than the Solar one, planets capable to host life can be found. Moreover, recent 
observations performed with the Atacama Large Millimeter Array (ALMA) of the Large and Small Magellanic Clouds (LMC and 
SMC, respectively), which have a metallicity of a factor ~3 and ~5 lower, respectively, than the Solar one, have revealed emission 
of complex organic molecules (COMs), i.e. organic species with more than 5 atoms. 
Methanol, \METH, was detected in star-forming regions associated with both galaxies (Shimonishi et al.~\citeyear{shimonishi18}; 
Sewi\l{}o et al.~\citeyear{sewilo18}), and methyl formate (HCOOCH$_3$) and dimethyl ether (CH$_3$OCH$_3$) 
were found in hot-cores of the LMC (Sewi\l{}o et al.~\citeyear{sewilo18}). Because these 
COMs are thought to be precursors of more complex biogenic molecules (see e.g. Caselli \& Ceccarelli~\citeyear{cec12}), 
these observational findings indicate that the basic bricks of organic chemistry can be found also in metal poor environments. 
These findings were reinforced by the recent detections of a hot molecular core, WB89-789, rich in COMs, in the extreme OG
(Shimonishi et al.~\citeyear{shimonishi21}), and of methanol in star-forming regions located at $R_{\rm GC}$ up to 
$\sim 23.5$~kpc (Bernal et al.~\citeyear{bernal21}).

Despite these recent findings, the formation and survival of molecules in star-forming regions located in the two environments 
(i.e. inner and outer Galaxy) is expected to be different. In fact, besides a lower initial abundance of heavy elements in the
OG, relative ratios of different elements change as well because their abudances do not vary in the same way 
as a function of $R_{\rm GC}$ (Esteban et al.~\citeyear{esteban17}). For example, the N/O ratio has a clear negative 
trend with $R_{\rm GC}$ (Magrini et al.~\citeyear{magrini18}). Moreover, \citet{berg16} suggest for the C/O ratio a 
flat relation at metallicity lower than the Solar one, and a steeply increasing slope at higher metallicities, i.e. with
decreasing $R_{\rm GC}$, even though different trends are found at low metallicities from state-of-the-art analysis 
of high-resolution spectra of solar neighbourhood halo dwarf stars (e.g.~Amarsi et al.~\citeyear{amarsi19}).
Both parameters, i.e. the initial elemental abundances and their relative ratios, are crucial inputs of chemical models that attempt
to reproduce the observed molecular emission in star-forming regions (e.g.~star-forming cores, Fontani et al.~\citeyear{fontani17}; 
protoplanetary disks, Eistrup et al.~\citeyear{eistrup18}; extragalactic environments, Sewi\l{}o et al.~\citeyear{sewilo18}). 
Therefore, to gain understanding on how chemistry changes throughout the Galaxy
and constrain models adapted for lower metallicity environments, it is crucial to observe OG 
star-forming regions in a large variety of molecular tracers.

With the aim of attacking this problem, we have started the observational project called 'CHEMical complexity 
in star-forming regions of the OUTer Galaxy' (CHEMOUT).
The immediate goal, based on observations with the IRAM 30m telescope, is to unveil the chemical composition in
star-forming regions of the OG. Thanks to these observations, we will start to investigate how molecules 
form in such low-metallicity environment. In particular, molecules that are potentially biogenic species,
such as COMs, carbon chains, and sulfur- and phosphorus-bearing molecules, are of particular interest since in 
star-forming cloud cores they are expected to be part of the planetesimals out of which planets form and/or can
be delivered to the surfaces of planets, where they might play key roles in the origin of life. For this reason, 
the results of CHEMOUT will also allow us to prepare the ground for a re-discussion of the concept of GHZ.
Moreover, the isotopic abundances and abundance ratios determined by CHEMOUT will be used to constrain 
Galactic chemical evolution and stellar nucleosynthesis models at low metallicities. CHEMOUT data will be 
especially useful to study the evolution of rare isotopes that lack spectroscopic determinations in stellar 
atmospheres - such as, for instance, $^{15}$N - or for which statistically relevant stellar samples are not 
in hand at present (e.g., $^{13}$C, $^{17}$O, and $^{18}$O).

In this first paper, we present the list of targets (Sect.~\ref{sample}), the first observational dataset (Sect.~\ref{observations}),
and show as first results the detection of several organic species and tracers of star-formation activity 
(such as SO and SiO). These results are shown and described in Sect.~\ref{res}, and discussed in Sect.~\ref{discu}. 
The main conclusions and the outlook of the project are given in Sect.~\ref{conc}.

\section{Sample}
\label{sample}

\begin{table}
\begin{center}
\caption{Spectral windows observed and observational parameters}
 \begin{tabular}{lccc}
\hline
Spectral  & HPBW   &  $V_{\rm res}$$^{(a)}$ &   $\eta_{\rm MB}$$^{(b)}$ \\
windows  &               &                   &                              \\
(GHz)      &   (\asec) &    (\kms )    &                              \\
\hline
85.310 - 87.130   & 28  &  $\sim 0.16$ & 0.85   \\
88.590 - 90.410    & 27  &  $\sim 0.16$ & 0.84   \\
151.750 -  153.570    & 15  &  $\sim 0.096$ & 0.77   \\
148.470 - 150.290     &  15 &  $\sim 0.096$ & 0.77   \\
\hline
\end{tabular}
\end{center}
$^{(a)}$ velocity resolution in the spectrum; \\
$^{(b)}$ defined as $\eta_{\rm MB}=B_{\rm eff}/F_{\rm eff}$, where $B_{\rm eff}$ is the main
beam efficiency and $F_{\rm eff}$ the forward efficiency. \\
\label{tab:obs}
\end{table}

We have observed 35 targets extracted from \citet{blair08}, who searched for formaldehyde emission 
with the Arizona Radio Observatory (ARO) 12m telescope in dense molecular cloud cores of the OG
associated with IRAS colours typical of star-forming regions.
The list of targets is given in Table~\ref{tab:sources}. 
We selected objects (1) clearly detected in H$_2$CO $J_{K_a,K_b}= 2_{1,2}-1_{1,1}$, and 
(2) covering as much as possible the full range of $R_{\rm GC}$ in the OG. The targets 
have $R_{\rm GC}$ in between $\sim 9$ and $\sim 24$~kpc from the Galactic Centre. 
Table~\ref{tab:sources} shows the target equatorial (J2000) coordinates. In the same Table
we also report two parameters taken from Blair et al.~(\citeyear{blair08}), namely the H$_2$ total 
column density, N(H$_2$), and the Galactocentric distance, $R^{old}_{\rm GC}$. 
In Sect.~\ref{dist} we will recompute $R_{\rm GC}$, $R^{new}_{\rm GC}$, based on the 
systemic velocity of the sources derived from our new data, and using an updated 
rotation curve of the Galaxy. For the objects included in the Hi-GAL catalogs 
(Molinari et al.~\citeyear{molinari16}), in Table~\ref{tab:sources} we also give 
the total gas mass ($M$) and bolometric luminosity ($L$) derived from {\em Herschel} 
observations (Elia et al.~\citeyear{elia21}) and rescaled for the new heliocentric distances 
calculated in this work (Sect.~\ref{dist}). 

\begin{figure}
{\includegraphics[width=9cm]{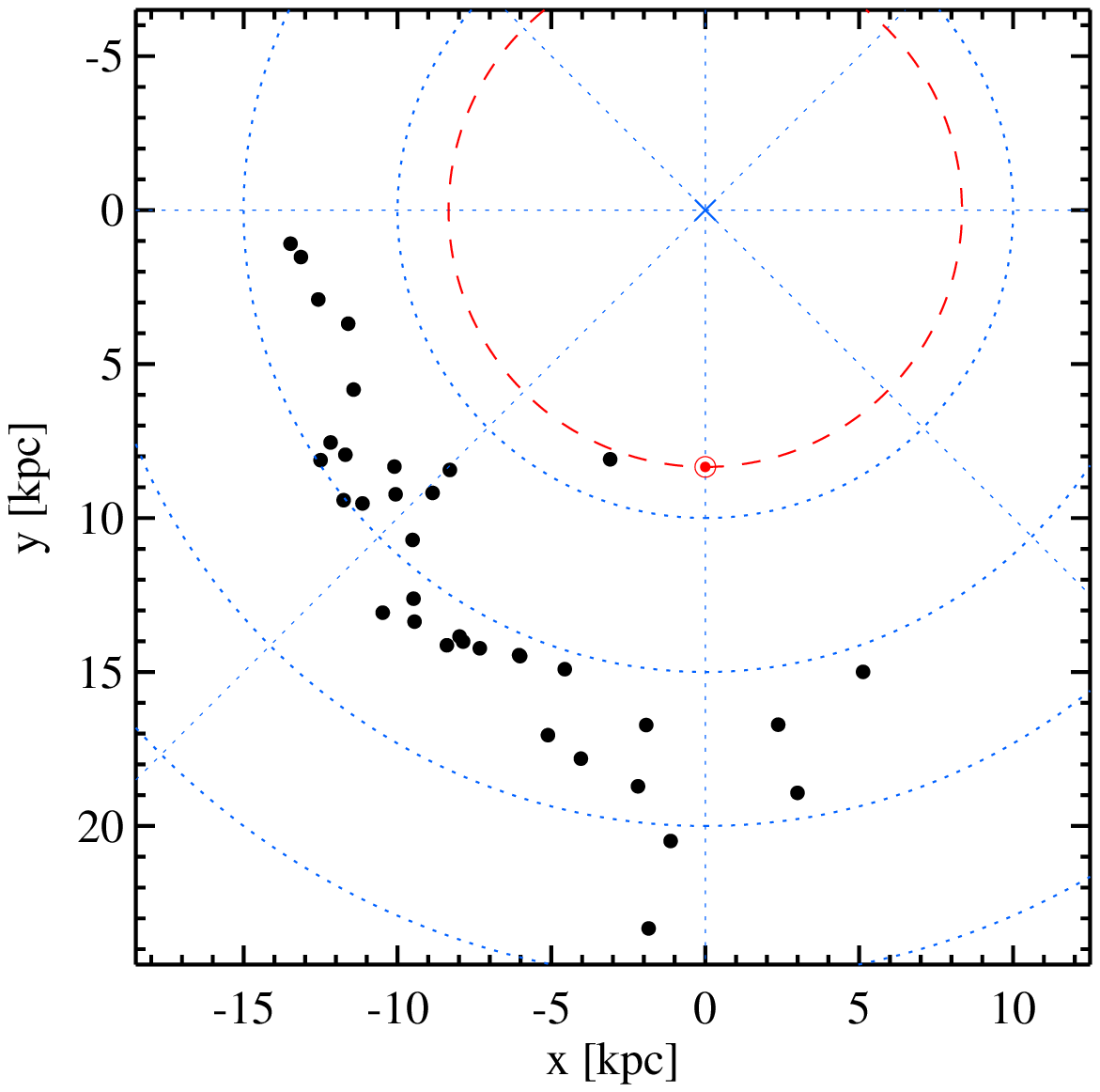}}
\caption{Plan view of the Galactic disk showing the location of the sources in Table~\ref{tab:sources}
(filled points) after recalculating their Galactocentric and heliocentric distances (Sect.~\ref{dist}). 
The Sun is marked by the red-circled point, and the Galactic centre is pinpointed by the
blue cross. The Solar circle is shown by the red dashed curve. The blue dotted circles correspond to
Galactocentric distances of 10, 15, 20, and 25~kpc.}
\label{fig:plane}
\end{figure}

The masses listed in Table~\ref{tab:sources} indicate that about half of the targets are 
candidate high-mass star-forming regions. Because the sources for which the mass has not been 
estimated yet have comparable distances and intensities of the \FORM\ lines (Blair et al.~\citeyear{blair08}), 
it is expected that all objects have similar physical properties, and hence they are all good candidate
high-mass star-forming regions. However, only accurate measurements of their gas masses will 
allow to confirm (or deny) this hypothesis.
The location of the sources in the Galactic plane is illustrated in Fig.~\ref{fig:plane}. Most of the
sources are included in the II quadrant of the Galaxy. WB89-002, the only object in 
\citet{blair08} located very close to the Sun ($R_{\rm GC}\sim 8.6$~kpc), was included in this study to have
at least one target representative of the Sun neighbourhoods observed with exactly the same
observational setup as the OG sources. This will allow us to derive observational trends or gradients 
(e.g. abundance gradients or isotopic fraction gradients) from the OG to the Sun location without
extrapolating.

\section{Observations}
\label{observations}

\begin{table*}
\begin{center}
\setlength{\tabcolsep}{2pt}
\label{tab:detections}
\caption{Molecular lines of species detected in the spectral windows at 3~mm listed in Table~\ref{tab:obs} 
at a significance level well above 3$\sigma$ rms (Y) or comparable to 3$\sigma$ rms (Y?). }
 \begin{tabular}{lcccccccccccccc}
\hline
source     & rms$^{(1)}$ & \multicolumn{9}{c}{carbon-bearing species} & & \multicolumn{3}{c}{other species$^{(3)}$} \\
                &  (mK)       &                       &                  &                        &               &           &           &                              &         &                  & &                 &       &      \\
                \cline{3-11} \cline{13-15} \\
                &        & {\it c-}C$_3$H$_2$ & HCS$^+$ & CH$_3$CCH  & C$_4$H & CCS  & HCO  & H$^{13}$CO$^+$ & HCN & HCO$^+$$^{(2)}$ & & NH$_2$D & SO & SiO \\
\hline
WB89-315  & $\sim 9.1$  & n  & n  & n & n  & n  & n  & n & Y & Y(n)     & &  n & n &  n  \\
WB89-379  & $\sim 6.0$  & Y  & Y  & Y & n  & n  & Y  & Y  & Y & Y(y?)   & &  n & Y  &  Y?  \\
WB89-380  & $\sim 17$   & Y  & n  & Y? & n  & n  & Y  & Y  & Y & Y(2,y) & &  n & Y  &  Y   \\
WB89-391  & $\sim 6.6$  & Y  & Y? & n & n  & n  & Y  & Y  & Y & Y(y?)   & &  n & Y  &  Y   \\
WB89-399  & $\sim 9.3$  & Y  & n  & n & Y? & n  & Y  & Y? & Y & Y(y?)   & &  n & n &  Y?   \\
WB89-437  & $\sim 6.0$  & Y  & Y  & Y & Y  & Y? & Y  & Y  & Y & Y(y)    & & Y  & Y  &  Y   \\
WB89-440  & $\sim 6.7$  & Y  & n & n  & n  & n  & Y  & n & Y & Y(n)     & & n & n &  Y   \\
WB89-501  & $\sim 7.2$  & Y  & n  & n & n  & n  & Y  & Y  & Y & Y(y?)   & &  n & Y  &  Y   \\
WB89-529  & $\sim 11.5$ & Y  & n  & n & n  & n & Y  & n & Y & Y(n)     & &  n & n &  n   \\
WB89-572  & $\sim 26$   & Y  & n  & n & n  & n  & n  & Y? & Y & Y(y)    & &  n & Y &  n  \\
WB89-621  & $\sim 5.7$  & Y  & Y  & Y & n  & Y? & Y  & Y  & Y & Y(y)    & & n & Y  &  Y   \\
WB89-640  & $\sim 10.7$ & Y  & Y? & n & n  & n  & Y  & Y  & Y & Y(y)    & &  Y  & Y  &  n  \\
WB89-670  & $\sim 9.2$  & Y  & n  & n & Y  & Y  & n  & Y  & Y & Y(y)    & & n & n &  n \\
WB89-705  & $\sim 7.8$  & Y  & n  & n &  n  & n  & Y? & Y  & Y & Y(n)   & &  Y  & Y  &  n  \\
WB89-789  & $\sim 7.2$  & Y  & Y  & Y? & n  & n  & Y  & Y  & Y & Y(y)   & & Y  & Y  &  n  \\
WB89-793  & $\sim 13.2$ & Y  & n  & n & n  & n  & n  & Y  & Y & Y(y)    & & n & Y  &  n  \\
WB89-898  & $\sim 8.4$  & Y  & n  & n & n  & n  & Y? & n & Y & Y(y?)    & &  n & Y  &  n  \\
19383+2711 & $\sim 6.6$ & Y  & Y  & n & Y  & n & Y  & Y  & Y & Y(2,n)   & & n & Y  &  Y   \\
19423+2541 & $\sim 7.6$ & Y  & n & n & Y? & n & Y  & Y  & Y & Y(y)      & & Y? & Y  &  Y   \\
19489+3030 & $\sim 6.5$ & Y  & Y? & n & n  & n & n & Y  & Y & Y(y)     & &  Y  & Y? &  Y?  \\
19571+3113 & $\sim 6.0$ & Y  & Y? & n & n  & n  & Y  & Y  & Y & Y(2,n)   & &  n & n &  Y   \\
20243+3853 & $\sim 5.0$ & Y  & Y  & n & Y? & n & Y  & Y  & Y & Y(y?)    & & n & Y  &  Y   \\
WB89-002   & $\sim 14$  & Y  & n  & n & n  & n  & Y  & Y?  & Y & Y(n)    & & n & n &  n  \\
WB89-006   & $\sim 7.3$ & Y  & Y  & n & Y  & n  & Y? & Y  & Y & Y(2,y)  & & Y  & Y  &  n  \\
WB89-014   & $\sim 8.5$ & Y  & n  & n & n  & n  & n  & n & Y & Y(n)      & & n & n &  n  \\
WB89-031   & $\sim 6.5$ & Y  & n  & n & n  & n  & Y  & n & Y & Y(y?)    & & n & n &  Y?  \\
WB89-035   & $\sim 7.3$ & Y  & n  & n & n  & n  & Y  & Y  & Y & Y(y)    & & n & Y  &  Y   \\
WB89-040   & $\sim 6.2$ & Y  & Y  & n & Y? & n  & Y  & Y  & Y & Y(y?)    & &  n & Y  &  n  \\
WB89-060   & $\sim 7.1$ & Y  & Y  & Y  & n  & n  & n  & Y  & Y & Y(2,y)  & & n & Y  &  Y   \\
WB89-076   & $\sim 7.5$ & Y  & Y  & n & Y & Y  & Y  & Y  & Y & Y(y)    & &  Y  & Y  &  n  \\
WB89-080   & $\sim 8.3$ & Y  & n  & n & n  & n  & Y  & Y  & Y & Y(2,y)  & & n & Y  &  n  \\
WB89-083   & $\sim 20$  & Y  & n  & n & n  & n  & n  & Y & Y & Y(n)     & & n & n &  n  \\
WB89-152   & $\sim 14$  & Y  & n  & n & n  & n  & n  & n & Y & Y(n)     & & n & n &  n  \\
WB89-283   & $\sim 7.5$ & Y  & n  & n & n  & n  & Y  & Y & Y & Y(y)    & & n & Y? &  n  \\
WB89-288   & $\sim 7.4$ & Y  & n  & n & n  & n  & Y  & n & Y & Y(y?)    & &  n & n &  n  \\
\hline
\end{tabular}
\end{center}
$^{(1)}$ 1$\sigma$ rms noise at 3~mm, calculated for a velocity resolution of $\sim 0.65$~\kms ; \\
$^{(2)}$ HCO$^+$ $J=1-0$ line profile can show two velocity features (2), and/or high velocity wings (y/n); \\
$^{(3)}$ We do not report the detections in the D, $^{13}$C and $^{15}$N isotopologues of HCN and HNC, which 
will be published in forthcoming papers (Colzi et al. in prep.; Fontani et al. in prep.). \\
\end{table*}

The observations were performed with the IRAM 30m telescope in several observing runs
(5 days in March and April, 2018, for a total of $\sim 20$ hours; August 14-21, 2018 for additional $\sim 70$ hours). 
In all runs we used the 3mm and 2mm receivers simultaneously. At 3~mm, the spectral windows 
were optimised to observe the $J= 1-0$ transitions of the four less abundant isotopologues of HCN and HNC,
namely H$^{15}$NC, HN$^{13}$C, H$^{13}$CN, and HC$^{15}$N. At 2~mm, we centred the bands 
on the DNC $J=2-1$ line. All these transitions will be analysed in forthcoming papers (Colzi et
al., in prep.; Fontani et al., in prep.). The Local Standard of Rest (LSR) velocities used to centre the 
receiver bands are listed in Table~\ref{tab:sources}. Table~\ref{tab:obs}
shows the spectral ranges observed in the two receivers, as well as some technical details of
the observations: the beam full width at half maximum (HPBW), the velocity resolution ($V_{\rm res}$), 
and the telescope efficiency ($\eta_{\rm MB}$) used to convert the spectra from antenna temperature
to main beam temperature units.
The observations were made in wobbler-switching mode with a wobbler throw of 240\asec. Pointing 
was checked (almost) every hour on nearby quasars, planets, or bright HII regions. Focus was 
checked on planets at the start of observations, and after sunset and sunrise. The data were 
calibrated with the chopper wheel technique (see Kutner \& Ulich~\citeyear{keu81}), with a calibration uncertainty 
of about $10\%$. The spectra were obtained with the fast Fourier transform spectrometers with the 
finest spectral resolution (FTS50), providing a channel width of 50 kHz. In this work, the calibrated spectra 
were fitted with the {\sc class} package of the {\sc gildas}\footnote{https://www.iram.fr/IRAMFR/GILDAS/} 
software using standard procedures. The spectral 1$\sigma$ rms noise, strongly source-dependent, 
is given in Table~\ref{tab:detections}, and goes generally from $\sim 5$ to $\sim 26$~mK in the 3~mm band.
The 2~mm band, not presented in this work, will be shown in a forthcoming paper discussing
DNC emission (Fontani et al., in prep.).

\section{Results}
\label{res}

\begin{table*}
\begin{center}
\caption{Rest frequencies and other spectroscopic parameters of the detected molecular transitions.}
 \begin{tabular}{lcccc}
\hline
molecule   & rest & quantum & $E_{\rm u}$ & Log[$A_{i,j}$/s] \\
                  & frequency        & numbers  &                    &              \\
                  &   MHz               &                 &    K            &    \\
                  \hline
{\it c-}C$_3$H$_2$ & 85338.89 & $J_{K_a,K_b}=2_{1,2}-1_{0,1}$ &  6.4 & 	--4.6341 \\
\hline
HCS$^+$  & 85347.89 & $J=2-1$  &  6.1 & --4.9548 \\
\hline
CH$_3$CCH & 85457.30 & $J(K)=5(0)-4(0)$  & 12.3 & --6.20908 \\
                      & 85455.67 & $J(K)=5(1)-4(1)$  & 19.5 & --6.2268  \\
\hline
C$_4$H   & 85634.00 &  $N=9-8$, $J=19/2-17/2$, $F=9-8$  &  20.5   & --4.8189 \\
                & 85634.02  &  $N=9-8$, $J=19/2-17/2$, $F=10-9$ & 20.5  & --4.8163 \\
                & 85672.58  &  $N=9-8$, $J=17/2-15/2$, $F=8-7$   & 20.6  & --4.8217 \\
                & 85672.58  &  $N=9-8$, $J=17/2-15/2$, $F=9-8$   & 20.6  & --4.8184 \\
\hline
NH$_2$D & 85926.28 & $J_{K_a,K_b}=1_{1,1}-1_{0,1}$  & 20.7  & --5.1057 \\
\hline
CCS         & 86181.39 &  $N=7-6$, $J=6-5$ & 23.3  & 	--4.5563 \\
\hline
SO            & 86093.95    & $N=2-1$, $J=2-1$ & 19.3  & --5.2799 \\
\hline  
HCO         & 86670.76          &  $N_{K_a,K_b}=1_{0,1}-0_{0,0}$, $J=3/2-1/2$, $F=2-1$     & 4.2 & --5.3289  \\
                 & 86708.36          &  $N_{K_a,K_b}=1_{0,1}-0_{0,0}$, $J=3/2-1/2$, $F=1-0$     & 4.2 & --5.3377  \\ 
                 & 86777.46          &  $N_{K_a,K_b}=1_{0,1}-0_{0,0}$, $J=1/2-1/2$, $F=1-1$     & 4.2 &  --5.3366  \\
                 & 86805.78          &  $N_{K_a,K_b}=1_{0,1}-0_{0,0}$, $J=1/2-1/2$, $F=0-1$     & 4.2 & --5.3268   \\
 \hline
 H$^{13}$CO$^+$ & 86754.29 & $J=1-0$  & 4.2 & --4.4142  \\
 \hline  
 HCN & 88630.42  & $J=1-0$, $F=1-1$ & 4.3 & --4.6184 \\
          & 88631.85  & $J=1-0$, $F=2-1$ & 4.3 & --4.6185 \\
          & 88633.94  & $J=1-0$, $F=0-1$ & 4.3 & --4.6184 \\
 \hline 
 HCO$^+$ & 89.18852 & $J=1-0$ &  4.3 & --4.3781 \\
 \hline
\end{tabular}
\end{center}
\label{tab:spectroscopy}
\end{table*}

\subsection{Detection summary}
\label{summary}

Table~\ref{tab:detections} lists the species that have been detected at a significance level of $\geq 3\sigma$ rms 
noise in the 3~mm band (Table~\ref{tab:obs}), except for the D, $^{13}$C and $^{15}$N isotopologues of HCN 
and HNC, which will be listed and analysed in forthcoming papers (Colzi et al. in prep.; Fontani et al., in prep.). 
Rest frequencies, quantum numbers, energy of the upper level, and Einstein coefficients
of the detected transitions are listed in Table~\ref{tab:spectroscopy}, and are taken from the Cologne 
Database for Molecular Spectroscopy (CDMS\footnote{https://cdms.astro.uni-koeln.de/classic/}, 
Endres et al.~\citeyear{endres16}) and the Jet Propulsion Laboratory (JPL\footnote{https://spec.jpl.nasa.gov/ftp/pub/catalog/catdir.html}, 
Pickett et al.~\citeyear{pickett98}).
For the C$_4$H radical, we observed two doublets with quantum numbers $N= 9-8$, $J=19/2-17/2$ 
with $F= 9-8$, $10-9$, and $J=17/2-15/2$ with $F=8-7$, $9-8$.
For HCO we observed the quadruplet with quantum numbers $N_{K_a,K_b}=1_{0,1}-0_{0,0}$, 
$J=3/2-1/2$ with $F=2-1$, $1-0$, and $J=1/2-1/2$ with $F=1-1$, $0-1$ 
(see Table~\ref{tab:spectroscopy} for the rest frequencies and other spectroscopic 
parameters). The hyperfine components with different $F$ of C$_4$H cannot be resolved 
due to their negligible separation in frequency/velocity (Table~\ref{tab:spectroscopy}). 
This implies that the multiplet is grouped in two lines with different $J$ resolved in frequency/velocity.
Because these two lines have similar Einstein coefficients, we have considered the transitions 
as clearly detected when both lines with different $J$ have peak intensity $\geq 3\sigma$ rms, 
and as tentatively detected those for which one line is clearly detected and another one is tentatively 
detected. When the second one is not clearly detected, the species is not considered as detected 
even if the first one has an intensity higher than $ 3\sigma$ rms.

In Fig.~\ref{fig:wb89-437} we show the full 3~mm spectrum obtained towards the representative 
source WB89-437. All species and lines listed in Table~\ref{tab:detections}, and detected in the source,
are indicated in the plot. 
Selected spectral windows around the faintest detected lines are shown in Fig.~\ref{fig:wb89-437-zoom}.
We also give the line intensities of all detected lines for this source, bearing in mind that a thorough 
analysis of all species and lines goes beyond the scope of this presentation work, and will be 
performed in forthcoming papers.

We report the following detection rates: $35/35=100\%$ in HCN and \HCOp;
$34/35\sim 97\%$ in {\it c-}C$_3$H$_2$;  $27/35\sim 77\%$ in \HCOpI ; $25/35\sim 71\%$ in HCO;
$23/35\sim 66\%$ in SO; $16/35\sim 46\%$ in SiO; $14/35\sim 40\%$ in HCS$^+$; 
$9/35\sim 26\%$ in C$_4$H; $8/35\sim 23\%$ in NH$_2$D; $6/35\sim 17\%$ in CH$_3$CCH; 
$4/35\sim 11\%$ in CCS.
The detection rates of all species are listed in Table~\ref{tab:rates}.
The molecules with the largest detection rates, i.e. \HCOp, HCN, {\it c-}C$_3$H$_2$, and \HCOpI, 
are among the most abundant C-bearing species in the inner and local Galaxy (e.g.~Gozde et
al.~\citeyear{gozde18}; Gerner et al.~\citeyear{gerner14}; Kim et al.~\citeyear{kim20}), 
and their high detection rate indicates that these species are very abundant also in the OG.
The relatively high detection rate of HCO, together with that of {\it c-}C$_3$H$_2$, both
believed to be tracers of photodissociation regions in massive cores (Kim et al.~\citeyear{kim20}),
suggest that we are tracing likely the most extended envelope of the cores, 

In addition to simple organics, we report four clear detections and two tentative detections of the
complex hydrocarbon CH$_3$CCH in the $J=5-4$ rotational transition. The clear detections are 
obtained towards WB89-060, WB89-379, WB89-437, and WB89-621, while the tentative detections 
towards WB89-380 and WB89-789 (Table~\ref{tab:detections}). In particular, WB89-789 is found to
harbour a hot molecular core (Shimonishi et al.~\citeyear{shimonishi21}) rich in COMs and clearly detected
with ALMA in molecules even more complex than CH$_3$CCH, such as HCOOCH$_3$, C$_2$H$_5$OH, 
and CH$_3$OCH$_3$. Hence, we consider this tentative detection reliable. CH$_3$CCH is a known 
good thermometer of the intermediate density and temperature gaseous envelope of star-forming
regions (e.g. Giannetti et al.~\citeyear{giannetti17}). However, only the $K=0$ and 1 transitions are 
detected above the 3$\sigma$ rms level and hence cannot be used to constrain accurately the gas temperature.

\subsection{Occurrence of high-velocity wings in the \HCOp\ (1--0) line profile.}
\label{outflows}

To investigate the presence of on-going star-formation activity in our targets, we have searched 
for high-velocity emission in the wings of the HCO$^+$ $J=1-0$ line profiles, successfully used 
as tracers of protostellar outflows even at the large linear scales 
probed by our observations ($\sim 0.6-1.8$~pc). In fact, even though protostellar cores have sizes 
typically of $\sim 0.1$~pc, protostellar outflows are observed to be extended up to a few parsecs 
in \HCOp\ (e.g.~L\'opez-Sepulcre et al.~\citeyear{lopez10}; S\'anchez-Monge et al.~\citeyear{sanchez13}). 
On the other hand, as discussed in L\'opez-Sepulcre et al.~(\citeyear{lopez10}), 
the \HCOp\ line wings trace the outflows closer to the sources driving them with respect
to more abundant molecules (e.g. CO), which are more sensitive to the outer, lower density material.
Another typical outflow tracer even at pc-scales is SiO (e.g.~L\'opez-Sepulcre et al.~\citeyear{lopez11}), 
but in our data the SiO lines are detected in a smaller number 
of sources and with signal-to-noise ratio worse than in \HCOp. Therefore,
we first searched for outflows using the \HCOp\ lines, and then tried to check/confirm their presence 
in the SiO lines.

High-velocity wings in the HCO$^+$ $J=1-0$ lines were identified fitting the lines with Gaussian
profiles (one single Gaussian for lines with a single intensity peak, double Gaussians for lines 
with two peaks) and searched for deviations from the Gaussian shape in the high-velocity red- 
and blue-shifted wings. Fig.~\ref{fig:HCOp-Fig1} shows all spectra superimposed on the best
Gaussian fits, in which several high-velocity non-Gaussian wings are clearly identified. In 
Table~\ref{tab:detections} we report all sources that have evidence of high-velocity wings.
The line parameters obtained from the Gaussian fits are given in Appendix-B.
Some spectra also show multiple velocity features (WB89-380, 19383+2711, 19571+3113, WB89-006,
WB89-060, WB89-080). In this paper, we use this line only for the purpose of establishing 
the presence (or absence) of high-velocity wings. An extensive analysis of the \HCOp\ physical
parameters, and of the outflow properties eventually associated to the sources with high-velocity wings,
is beyond the scope of this paper and will be performed in a forthcoming paper.

We detect high-velocity wings in 16 targets, and tentatively towards 9 targets, for a total of 25 sources 
(i.e. $\sim 71\%$) likely associated with molecular outflows. We defined as tentatively detected a 
non-Gaussian high-velocity wing if the excess emission with respect to the Gaussian fit is detected 
at $3 - 6$ times the $1\sigma$ rms level. If the excess emission is above $6\sigma$, 
the detection is considered as firm. Inspection of Table~\ref{tab:detections} indicates that most 
of the sources associated with high-velocity wings in HCO$^+$ $J=1-0$ are detected in both SiO and SO. 
In particular, 13 out of the 16 sources detected in SiO are associated with \HCOp\ high-velocity
wings ($\sim 82\%$), and 21 out of the 23 sources detected in SO are associated with \HCOp\ 
high-velocity wings ($\sim 91\%$). This clearly shows that the SiO and SO lines are powerful 
probes of protostellar outflow activity also in the lower metallicity environment of the OG. 
Vice-versa, only 13 out of the 25 targets ($\sim 52\%$) associated with \HCOp\ high-velocity 
wings are detected in SiO, and 21 out of 25 in SO ($\sim 84\%$). However, this statistical difference 
can be due to the fact that the SiO lines are on average less intense than the SO ones. In fact, the
sources that show clear SiO emission tend to be associated with more intense HCO$^+$ $J=1-0$ 
lines (average peak temperature of HCO$^+$ $J=1-0$ $\sim 1.4$~K and $\sim 0.9$~K in sources 
detected and undetected in SiO, respectively; compare Table~\ref{tab:detections} 
and Fig.~\ref{fig:HCOp-Fig1}).

\subsection{Derivation of updated kinematic distances}
\label{dist}

We estimate new kinematic Galactocentric distances of the sources based on the velocities along 
the line of sight derived from the emission of {\it c-}C$_3$H$_2$ $J_{K_a,K_b}=2_{1,2}-1_{0,1}$. This 
latter is assumed to be optically thin based on the expected low abundance of the species. 
The spectra used are shown in Fig.~\ref{fig:C3H2-Fig1}. The lines are well fitted by a
Gaussian profile in almost all sources, confirming that the assumption of optically thin
emission is very likely satisfied. For two sources, 19571+3113 and 19383+2711, the {\it c-}C$_3$H$_2$
line shows two intensity peaks, which could be due to multiple velocity components or
to (self-)absorption. Because two velocity features at the same peak velocities 
are detected also in other molecular lines, we consider the two peaks as being due to
two gaseous clumps at sightly different velocity along the line of sight, and compute 
the kinematic distance for both.

For the sources that show a single peak in the spectrum of {\it c-}C$_3$H$_2$ and two peaks in
that of \HCOp, i.e. WB89-006, WB89-380, WB89-060, and WB89-080, this difference may 
be due either to the non-detection of the second, fainter velocity
feature in {\it c-}C$_3$H$_2$, or to (self-)absorption in the \HCOp\ line (not present in the
optically thin {\it c-}C$_3$H$_2$ line). In WB89-006 and WB89-380, the second option is the most
likely one, because the peak of the {\it c-}C$_3$H$_2$ line falls in between the two peaks detected
in \HCOp\ (compare Figs.~\ref{fig:HCOp-Fig1} and \ref{fig:C3H2-Fig1}). On the other hand, for 
WB89-060 and WB89-080 the strongest peaks in both molecules coincide in velocity, and the 
second velocity feature seen in \HCOp\ but much fainter than the main one, is likely under the noise in 
{\it c-}C$_3$H$_2$.

We use the revisited rotation curve of the Galaxy of Russeil et al.~(\citeyear{russeil17}), given by 
$\Theta(R_{\rm GC})/\Theta_{\odot} = 1.022 (R_{\rm GC}/R_{\odot})^{0.0803}$, where $\Theta(R_{\rm GC})$
is the rotation velocity at $R_{\rm GC}$, and $R_{\rm \odot} = 8.34$~kpc and $\Theta_{\odot}= 240$~\kms\ 
are the Galactocentric distance and rotation velocity of the Sun, respectively (Reid et al.~\citeyear{reid14}). 
The new $R_{\rm GC}$, $R^{new}_{\rm GC}$, and the parameters used to estimate them are given in 
Table~\ref{tab:newdist}. We find a good agreement between $R^{new}_{\rm GC}$ and $R^{old}_{\rm GC}$ 
(given in Table~\ref{tab:sources}), even though $R^{new}_{\rm GC}$ is always lower than 
$R^{old}_{\rm GC}$ by $\sim 0.1-1.5$ kpc (i.e., by $\sim 10\%$ at most). This is likely 
due to the fact that the previous estimates were based for most objects on the rotation curve of 
Brand \& Blitz~(\citeyear{beb93}) and on previous estimates of the parameters of the Solar motion 
$\Theta_{\odot}$ and $R_{\rm \odot}$. 

The uncertainty on $R^{new}_{\rm GC}$, calculated propagating 
the errors on $\Theta_{\odot}$, $R_{\rm \odot}$ and the line velocities, are of the order of $\sim 5-10\%$.
Table~\ref{tab:newdist} also lists the heliocentric distances, $d$, calculated from $R^{new}_{\rm GC}$.
They are in the range $\sim 8.0 - 15.3$~kpc, except for WB89--002, for which $d\sim 3.1$~kpc.
As also discussed in M\`ege et al.~(\citeyear{mege21}), the distances derived with this method can be
influenced by a true rotation pattern of the Galaxy more complicated than that of the assumed rotation curve.
Departures in velocity from circular rotation are typically of 10~\kms\ (Anderson et al.~\citeyear{anderson12};
Wienen et al.~\citeyear{wienen15}), but can be as high as 40 \kms\ (Brand \& Blitz~\citeyear{beb93}). Hence
these estimates can be associated with uncertainties of a few kpc.

We have derived kinematic distances also from the rotation curve of Reid et al.~(\citeyear{reid19}).
The values are given in Table~C-1 of Appendix-C. They are on average 
smaller than, but consistent with, those given in Table~\ref{tab:newdist} within a $\sim 20-30\%$ at 
most. The three largest differences are found in WB89-670, WB89-705, and WB89-002, for which 
from the curve of Reid et al.~(\citeyear{reid19}) $d$ is smaller by about a factor 2, and $R_{\rm GC}$ 
in WB89-670, WB89-705 is smaller by a factor 1.4. 
These are the nearest (WB89-002) and farthest (WB89-670 and WB89-705) targets of the sample, 
respectively. It is not straightforward to decide which rotation curve is most appropriate for the OG 
because both have not been sufficiently tested in this portion of the Milky Way. However, we decided 
to ultimately adopt the distances derived from the curve of Russeil et al.~(\citeyear{russeil17}) to be 
consistent with the approach used for the Hi-GAL catalogue. In fact, in our discussion we adopt some 
distance-dependent parameters given in Hi-GAL (Table~\ref{tab:sources}), which are computed, 
when possible, using the rotation curve of Russeil et al.~(\citeyear{russeil17}). Of course, when 
discussing possible observational Galactocentric trends, we will have to take with caution any trend 
particularly influenced by WB89-670, WB89-705, and WB89-002.

From the N(H$_2$) averaged within 44\asec\ (Blair et al.~\citeyear{blair08}, Table~\ref{tab:sources}), 
and taking the heliocentric distances into account (Table~\ref{tab:newdist}), we estimate that the 
average H$_2$ volume densities on such angular scale is of the order of $10^3$~\cmc.

\section{Discussion}
\label{discu}

\begin{table}
\begin{center}
\label{tab:rates}
\caption{Detection rate for the molecular lines detected in the spectral windows at 3~mm (Table~\ref{tab:obs}).}
 \begin{tabular}{lc}
\hline
species   & det. rate \\
\hline
HCN & 100$\%$ \\
HCO$^+$ & 100$\%$ \\ 
{\it c-}C$_3$H$_2$ & 97$\%$ \\
H$^{13}$CO$^+$ & 77$\%$ \\
HCO  & 71$\%$\\
SO & 66$\%$ \\
SiO & 46$\%$ \\
HCS$^+$ & 40$\%$ \\
C$_4$H & 26$\%$ \\
NH$_2$D & 23$\%$ \\
CH$_3$CCH  & 17$\%$ \\ 
CCS  & 11$\%$ \\
\hline
\end{tabular}
\end{center}
\label{tab:rates}
\end{table}

\begin{figure*}
{\includegraphics[width=9cm]{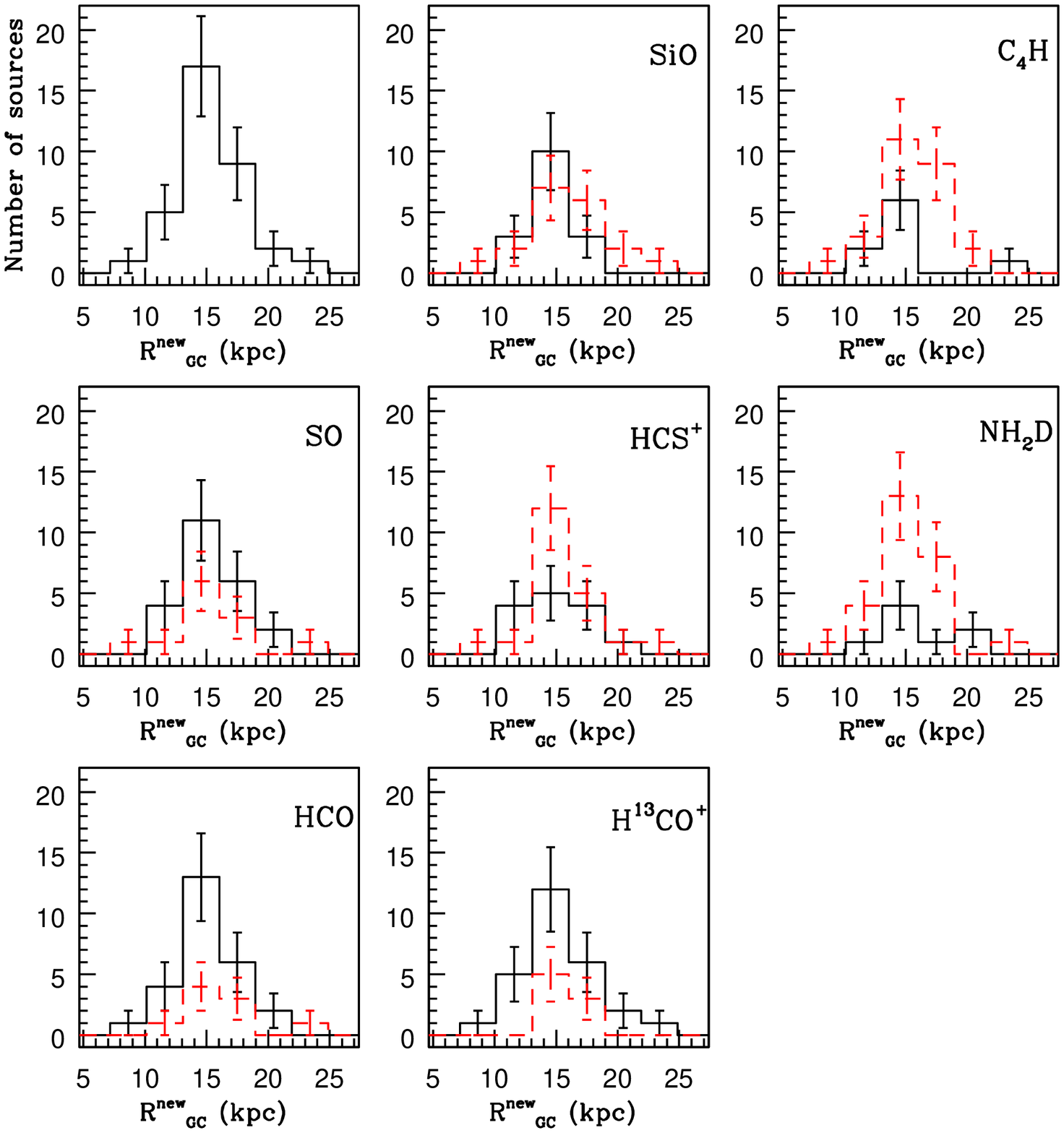}
\includegraphics[width=9cm]{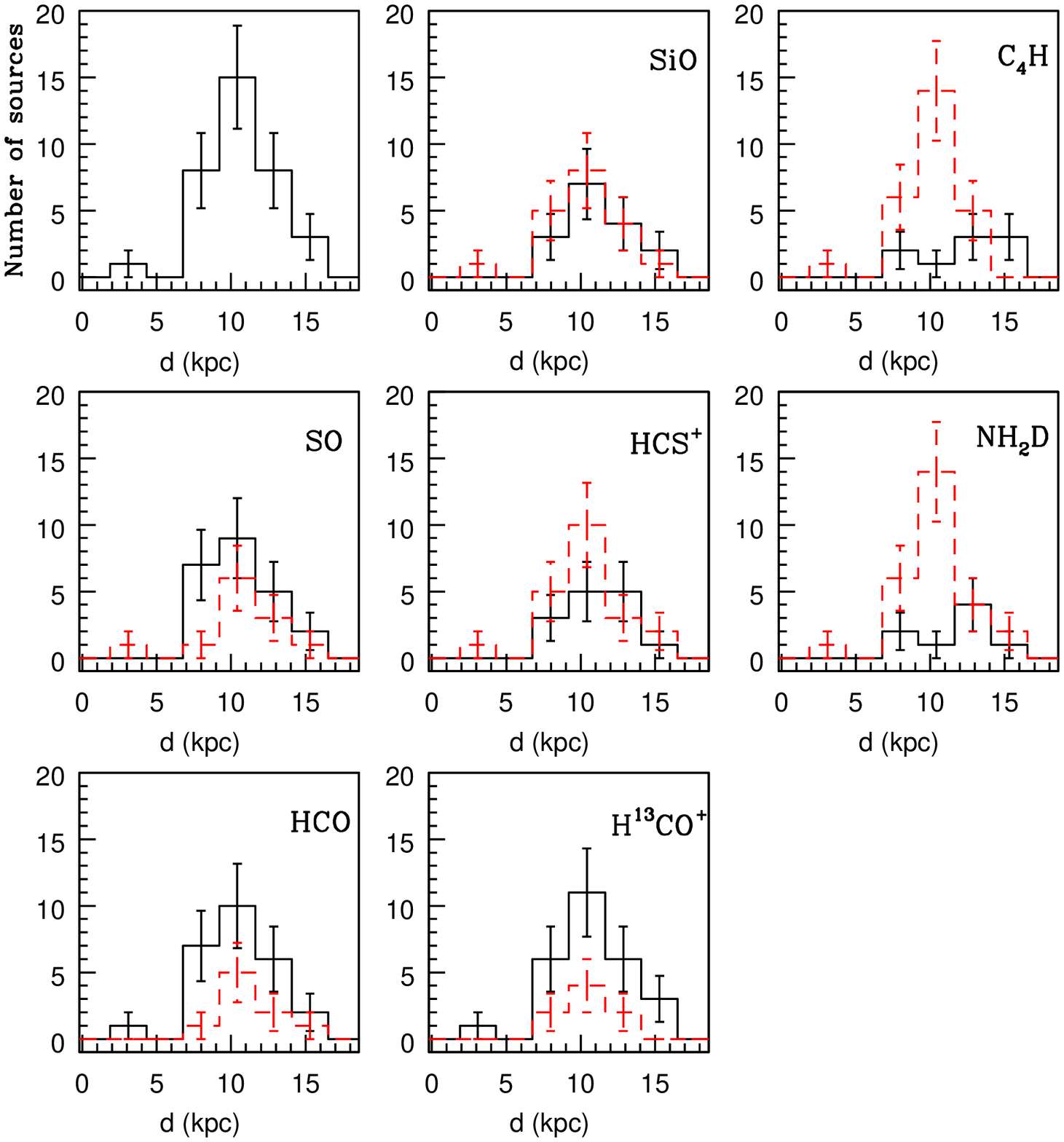}}
 \caption{Histograms comparing the Galactocentric (left panels) and heliocentric (right panels)
 distance for detected (solid line) and undetected (red-dashed line) sources in the molecular species 
 labelled in the top right corner of each panel. 
 The top left panel includes the whole sample without distinguishing between
 detections and non-detections. Poissonian errorbars are given in each bin. 
 }
\label{fig:statistics1}
\end{figure*}

\begin{figure*}
{\includegraphics[width=9cm]{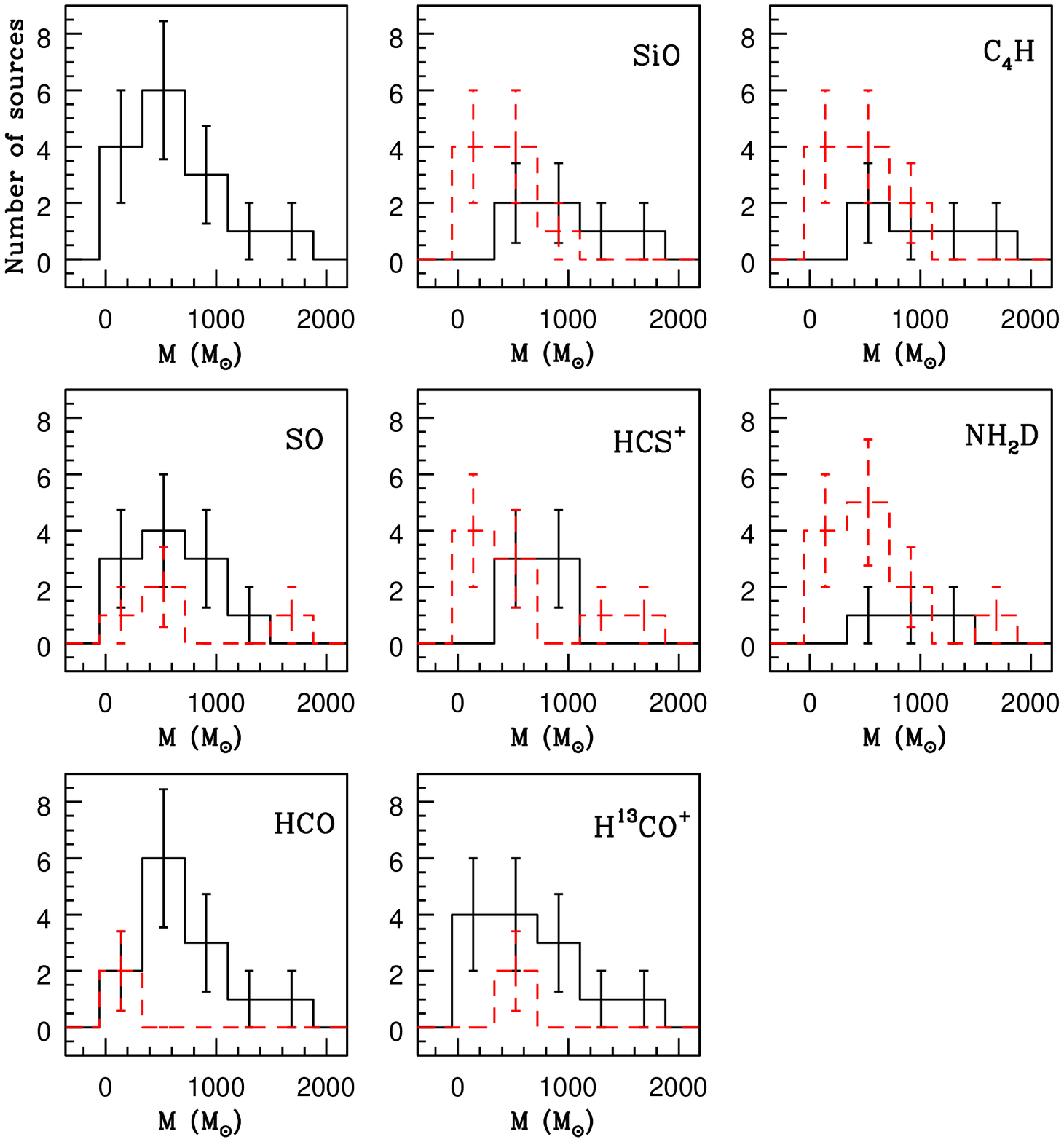}
\includegraphics[width=9cm]{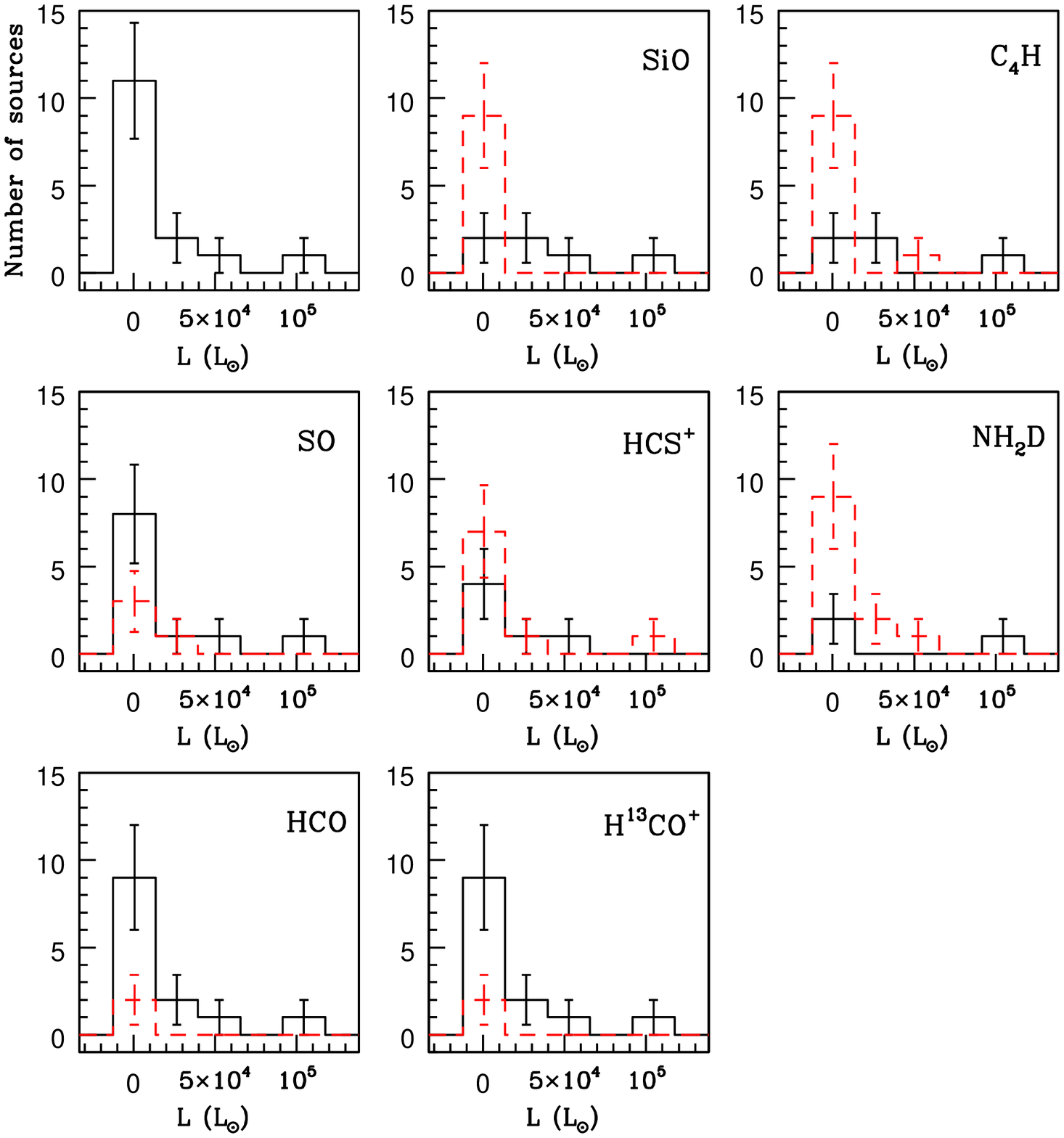}}
\caption{Same as Fig.~\ref{fig:statistics1} for H$_2$ mass ($M$, left panels), and the 
bolometric luminosity ($L$ right panels). Both parameters are derived from {\em Herschel}
observations and are given in Table~\ref{tab:sources}. }
\label{fig:statistics2}
\end{figure*}

\begin{figure*}
{\includegraphics[width=9cm]{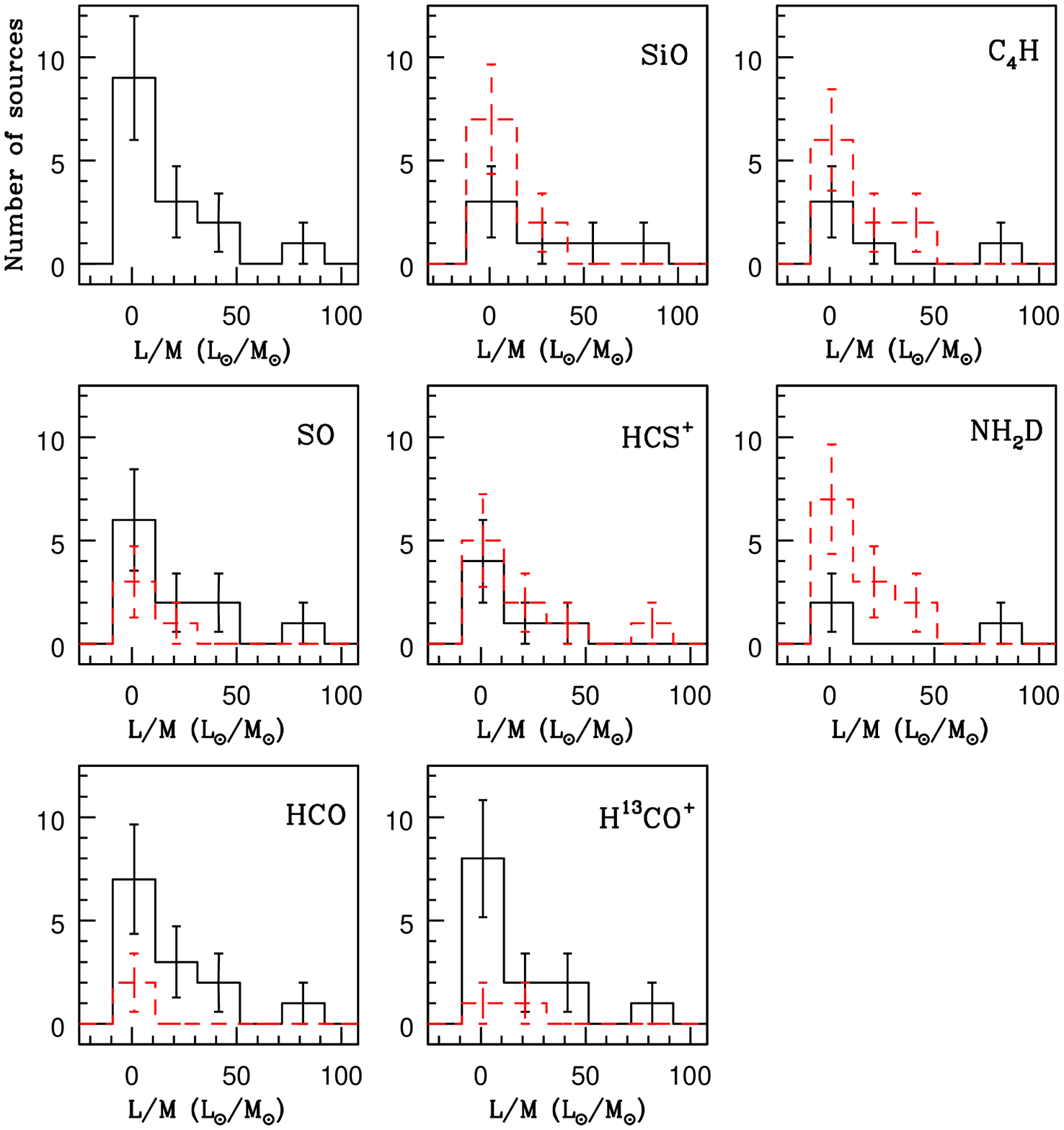}
\includegraphics[width=9cm]{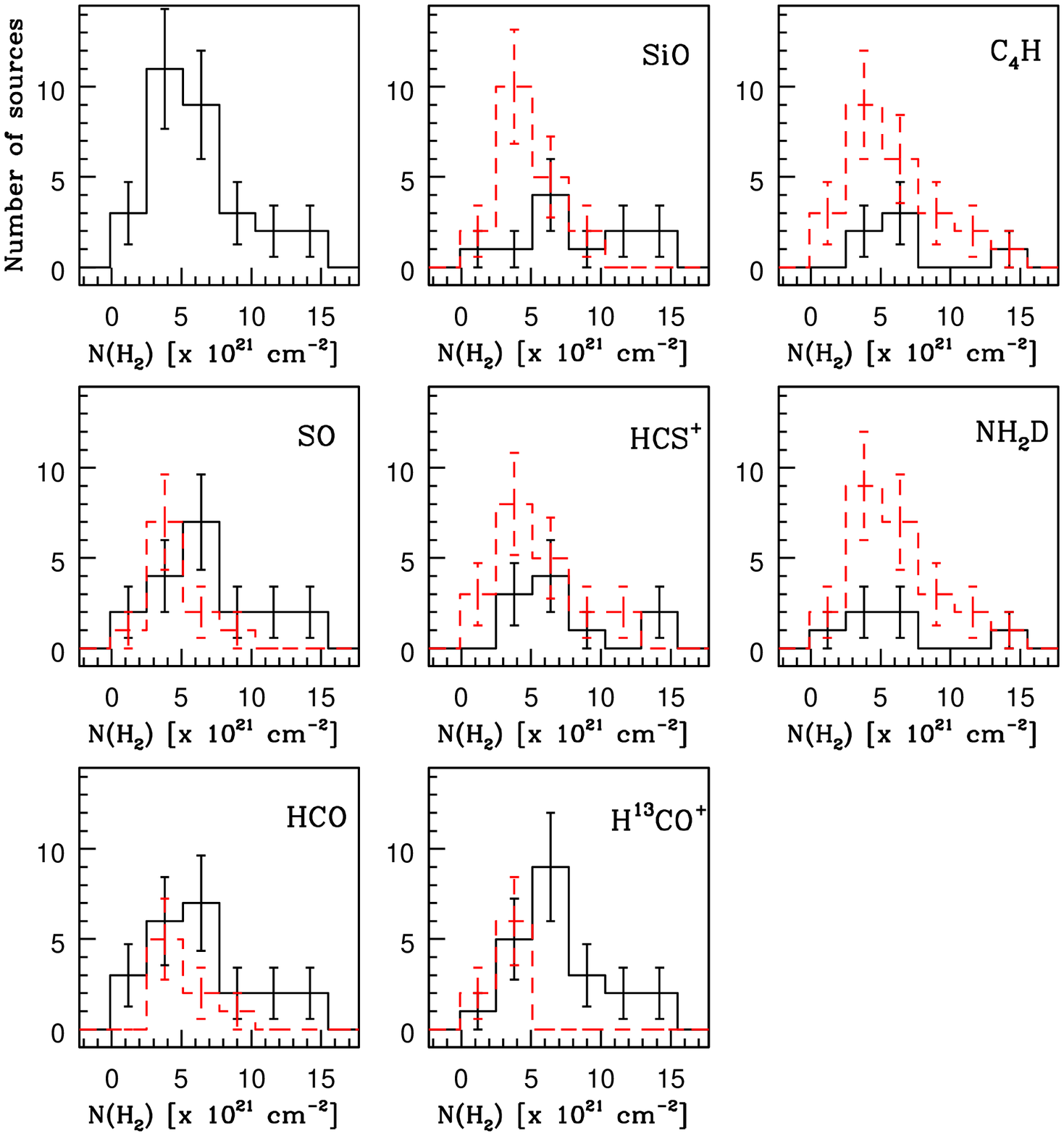}}
 \caption{Same as Fig.~\ref{fig:statistics2} for the luminosity-to-mass ratio ($L/M$, left
 panels) and the H$_2$ column density ($N({\rm H_2})$, right panels). The latter, derived by Blair et 
 al.~(\citeyear{blair08}), is given in Table~\ref{tab:sources}. }
\label{fig:statistics3}
\end{figure*}

\begin{figure}
{\includegraphics[width=9cm]{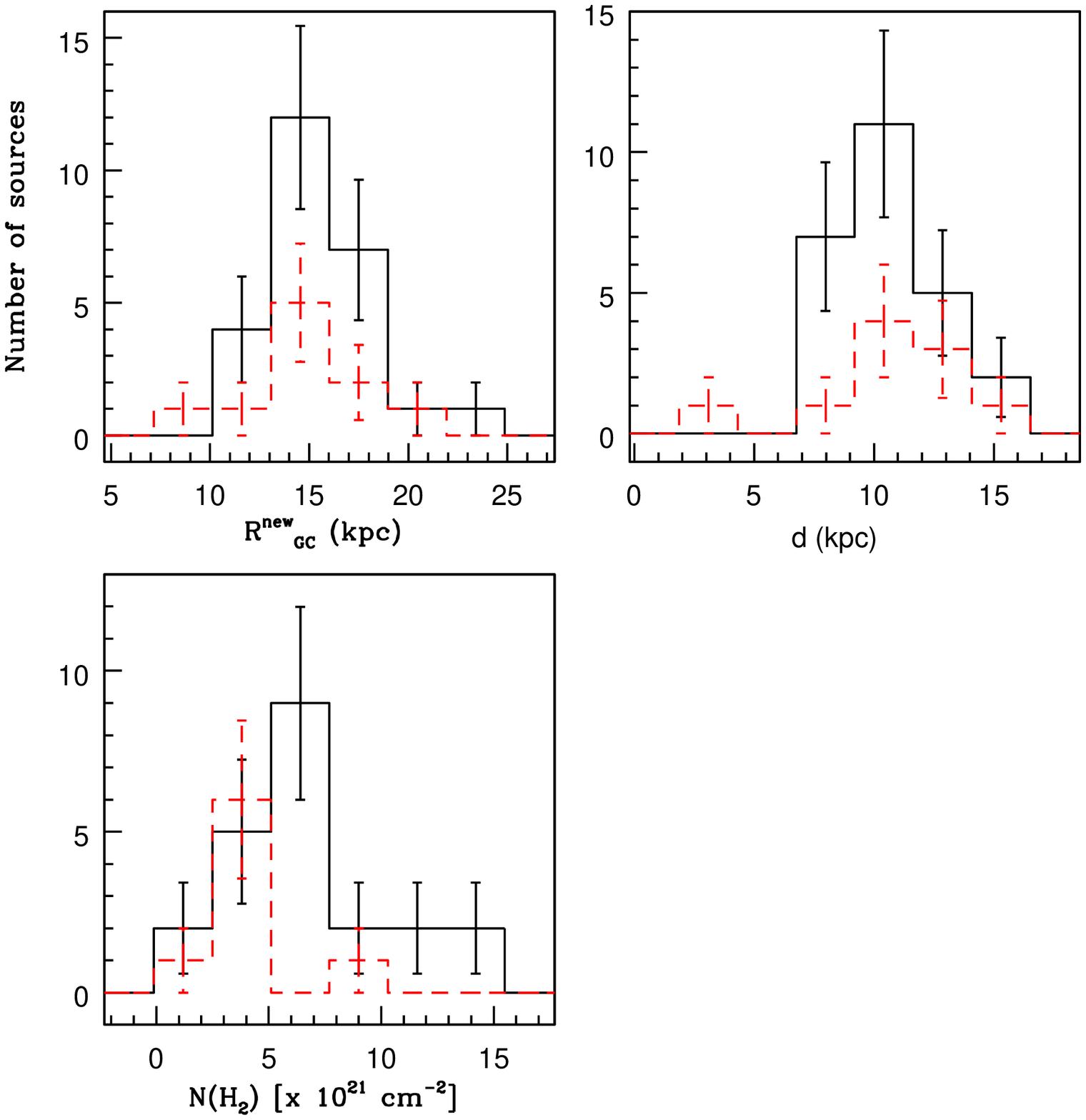}}
 \caption{Histograms showing the distribution of the sources associated
 (solid line) or not associated (red-dashed line) with \HCOp\ $J=1-0$ high-velocity
wings as a function of $R^{new}_{\rm GC}$, $d$, and $N({\rm H_2})$). Sources tentatively
associated with wings are included in the solid-line histograms.}
\label{fig:statisticswings}
\end{figure}

\subsection{Statistical analysis}
\label{analysis}

Here we analyse statistically the detection of the various molecular species 
as a function of the physical properties of the targets, i.e.: updated Galactocentric 
and heliocentric distance (Table~\ref{tab:newdist}), H$_2$ column density (Table~\ref{tab:sources})
and, for the sources belonging to the Hi-GAL catalogue, mass, luminosity, and luminosity-to-mass ratio
(Table~\ref{tab:sources}).
A thorough analysis of each molecular species and of the parameters that can be derived from 
them (in particular column densities and abundances) goes beyond the scope of this presentation
paper and will be performed in forthcoming papers.

Figure~\ref{fig:statistics1} shows the histograms with the detected and undetected sources in
each molecular tracer as a function of $R_{\rm GC}$ (left panels) and $d$ (right panels). 
We include only molecular species for which the detection rate is in between 
$20 - 80 \%$, so that the statistical comparison between the two groups (detected and undetected 
sources) is possible. The distribution for detected and undetected sources is similar, and similar
also to the total one for all molecules. Marginal differences between the two distributions are
tentatively seen in C$_4$H and NH$_2$D, but the detected sources are less than $\sim 30\%$
in both molecules and hence the comparison must be very cautious here.

Figures~\ref{fig:statistics2} and \ref{fig:statistics3} investigate, for each species, statistically significant
differences for detected and undetected sources as a function of $M$, $L$, 
luminosity-to-mass ratio ($L/M$), and $N$(H$_2$). In particular, $L/M$ is often used
as an evolutionary tracer because expected to increase with time, as the envelope mass decreases
while the bolometric luminosity increases during collapse. Overall, we do not find clear differences
between the distribution of detected and undetected sources also in this case, even though
the sources with high $L$ and $N(\rm H_2)$ tend to be always detected.
We suggest a tentative, but interesting, difference in the plots showing the results for SiO:
the sources associated with SiO emission tend to have higher $M$, $L$, and $N(\rm H_2)$. 
This could be interpreted as the result of a more active star formation activity in the most luminous
and massive objects, as already suggested by L\'opez-Sepulcre et al.~(\citeyear{lopez11})
in 57 high-mass molecular clumps mostly located in the inner Galaxy.

In this respect, we also investigated if the presence of high-velocity blue- and red-shifted 
emission in the \HCOp\ $J=1-0$ line depends somehow on $R^{new}_{\rm GC}$, $d$, and/or $N(H_2)$.
Figure~\ref{fig:statisticswings} shows the comparative histograms which, overall, again
do not show a clear difference between the distribution of detected and undetected sources 
as a function of the mentioned parameters, indicating once more that protostellar
activity does not seem to depend on $R_{\rm GC}$. Very similar distributions are also
found for detected and undetected sources as a function of $N({\rm H_2})$, even though
the detected sources tend to be associated with sources with higher $N({\rm H_2})$,
which can be due to sensitivity reasons: the high-velocity wings are detected more easily
towards more massive outflows.
As for possible differences between sources with or without wings as a function
of $M$, $L$, and $L/M$, unfortunately, only two sources with available $M$ and $L$ 
are undetected in \HCOp\ high-velocity wings (WB89-083 and WB89-529), hence a statistical 
comparison cannot be performed. 

\subsection{Towards a redefinition of the Galactic Habitable Zone?}
\label{ghz}

Our study clearly supports previous claims that organic molecules and tracers of protostellar 
activity are ubiquitous in the Galaxy (Blair et al.~\citeyear{blair08}, Bernal et al.~\citeyear{bernal21}). 
As said in Sect.~\ref{intro}, the outermost edge of the Milky Way was believed to be a 
hostile environment for both planets and (biogenic) organic molecules because 
the metallicity, i.e. the abundance of elements heavier than helium, is too low. Hence, the OG 
was excluded from the GHZ. This latter was defined as the region
with the most favourable conditions for the development and long-term maintenance of complex life 
comparable to terrestrial animals and complex plants (Gonzalez et al.~\citeyear{gonzalez01}). 

Based on this definition, two properties were invoked: (1) a sufficient amount of heavy elements to
form rocky planets and the basic bricks of biogenic molecules, and (2) a low concentration of high 
energy, potentially disruptive events such as supernovae and gamma-ray bursts, which may
cause life perturbation and extinction. Property (1) sets the outer boundaries of the GHZ, 
which depends on the metallicity gradient in the radial disk. Property (2)
sets the inner boundaries, as the concentration of potentially disruptive events would decrease 
with the distance from the centre of the Milky Way. Combining these two requirements, the GHZ of the
Milky Way was proposed to be an annulus of a few kpc centred at 8~kpc from the Galactic Centre 
(Spitoni et al.~\citeyear{spitoni17}), even though a hot debate is on-going on this subject.
E.g. Spinelli et al.~(\citeyear{spinelli21}) have proposed that the boundaries of the GHZ change
with Galaxy evolution, and that in the last four billion years the safer region to avoid 
disruptive events is $R_{\rm GC}\sim 2-8$~kpc, while the OG was safer for the development
of living organisms before that time.

Therefore, the boundaries of the GHZ are not rigid limits but rather based on probability
arguments, and the ubiquitous presence of rocky planets in the Galaxy, independent on the metallicity 
of the host environment (see Sect.~\ref{intro} and references therein), shows the need for a re-discussion, 
at least, of property (1). The idea of a redefinition of the GHZ is already discussed 
in \citet{blair08} and \citet{bernal21}, based on observations of single organic molecules (\FORM\ 
in Blair et al.~\citeyear{blair08}, \METH\ in Bernal et al.~\citeyear{bernal21}) in the OG. The 
CHEMOUT project, thanks to observations of multiple species and lines, obtained with the 
same setup(s) for all sources, place us in a position to expand this discussion by comparing 
in a consistent way the observational properties of many molecules (possibly) chemically connected.
In future papers, by comparing key observational parameters (in particular column densities 
and fractional abundances) with chemical models with adapted metallicity, we will better understand 
what are the main formation routes, and whether they are similar or different to those known to be 
efficient in the local and inner Galaxy. In this respect, the partial results presented in this work, and 
the similar recent observational findings of Bernal et al.~(\citeyear{bernal21}), suggest that the capacity 
of the environment to form organic (both simple and complex) molecules is independent on metallicity.
Thus, the outer boundaries of the GHZ are likely much wider than previously claimed 
because the basic bricks of organic chemistry can be easily found even at the outer edge 
of the Milky Way.

\section{Conclusions}
\label{conc}

\begin{table}
\begin{center}
\label{tab:newdist}
\caption{New Galactocentric ($R^{new}_{\rm GC}$) and heliocentric ($d$) distances computed from 
the peak velocity ($V_{\rm p}$) of the {\it c-}C$_3$H$_2$ $J_{K_a,K_b}=2_{1,2}-1_{0,1}$ line as 
explained in Sect.~\ref{dist}.}
 \begin{tabular}{lcccc}
\hline
source   & Longitude & $V_{\rm p}^{(a)}$ & $R^{new}_{\rm GC}$ & $d$ \\
              &     $\deg$      &     \kms\             &   kpc               & kpc \\
\hline
    WB89-315  &       118.0  &    --95.1(0.3)$^{(b)}$  &   16.3    & 10.7  \\
    WB89-379  &     124.6  &    --89.16(0.06)  &    16.4   &  10.2  \\
    WB89-380  &     124.6  &    --86.68(0.04)  &   16.0    & 9.7  \\
    WB89-391  &     125.8  &     --86.10(0.02)  &  16.1     & 9.7  \\
    WB89-399  &     128.8  &    --82.15(0.05)  &   16.0     & 9.4  \\
    WB89-437  &     135.3  &    --72.14(0.06)  &   15.7    & 8.6  \\
    WB89-440  &     135.6  &    --71.88(0.05)  &   15.7    &  8.6  \\
    WB89-501  &     145.2  &    --58.43(0.04)  &   15.6    & 8.0  \\
    WB89-529  &     149.6  &    --59.8(0.2)  &     17.8   &  10.1  \\
    WB89-572  &     156.9  &    --47.4(0.1)  &     18.3   &  10.3  \\
    WB89-621  &     168.1  &    --25.68(0.07)  &   18.9    & 10.6  \\
    WB89-640  &     167.1  &    --24.93(0.08)  &   16.8     & 8.6  \\
    WB89-670  &       173.0  &    --17.65(0.01)  & 23.4     & 15.1  \\
    WB89-705  &     174.7  &     --12.20(0.01)  &  20.5     & 12.2  \\
    WB89-789  &     195.8  &     34.25(0.05)  &    19.1   &  11.0  \\
    WB89-793  &     195.8  &     30.5(0.2)  &      16.9   &  8.7  \\
    WB89-898  &     217.6  &     63.5(0.1)  &      15.8   &  8.4  \\
  19423+2541  &     61.72  &    --72.58(0.04)  &    13.5   &  15.3  \\
  19383+2711  &     62.58  &    --70.2(0.2)  &   13.2    & 14.8  \\
  19383+2711-b$^{(c)}$ & 62.58 & --65.6(0.2) & 12.7   & 14.2 \\
  19489+3030  &     66.61  &    --69.29(0.05)  &   12.9    & 13.7  \\
  19571+3113  &     68.15  &    --61.7(0.1)  &     12.2   &  12.5  \\
  19571+3113-b$^{(c)}$ & 68.15 & --66.2(0.1) &   12.5   &  13.0 \\
  20243+3853  &      77.6  &    --73.21(0.05)  &   12.8    & 11.7  \\
    WB89-002  &     85.41  &     --2.83(0.09)  &   8.65   &  3.1  \\
    WB89-006  &     86.27  &    --90.38(0.05)  &    14.3    & 12.2  \\
    WB89-014  &     88.99  &    --96.0(0.1)  &     14.9   &  12.5  \\
    WB89-031  &     88.06  &    --88.89(0.08)  &   14.1    & 11.7  \\
    WB89-035  &     89.94  &    --77.56(0.03)  &   13.1    & 10.1  \\
    WB89-040  &     90.68  &    --62.38(0.05)  &   11.9    & 8.3  \\
    WB89-060  &     95.05  &    --83.7(0.15)  &    13.6   &  10.1  \\
    WB89-076  &     95.25  &    --97.07(0.02)  &   15.1    & 11.8  \\
    WB89-080  &     95.44  &    --74.1(0.2)  &     12.8   &  8.9  \\
    WB89-083  &     96.08  &    --93.76(0.04)  &   14.7    & 11.2  \\
    WB89-152  &       104.0  &    --88.5(0.2)  &   14.4    & 9.8  \\
    WB89-283  &     114.3  &    --94.69(0.06)  &   15.8    & 10.4  \\
    WB89-288  &     114.3  &      --101.0(0.1)  &  16.8    & 11.5   \\
    \hline
\end{tabular}
\end{center}
$^{(a)}$ obtained from the Gaussian fits in Table~A-2. Uncertainties derived from the fitting procedure
are quoted in parentheses; \\
$^{(b)}$ source undetected in {\it c-}C$_3$H$_2$ and in any other optically thin transition (Table~\ref{tab:detections}).
We use the $V_{\rm LSR}$ given in Blair et al.~(\citeyear{blair08}); \\
$^{(c)}$ second velocity feature (see Fig.~\ref{fig:C3H2-Fig1}). \\
\end{table}

With the IRAM-30m Telescope, we have searched for emission of simple organic molecules and
tracers of star-formation activity in 35 star-forming regions of the OG. Their (updated) 
Galactocentric distances are in between $\sim 8.7$ and $\sim 23.4$~kpc.
We report the detection of several simple organic molecules (HCN, \HCOp, \HCOpI, 
HCO, HCS$^+$, C$_4$H) and of the complex hydrocarbon CH$_3$CCH. We also detected
transitions of SO, CCS, NH$_2$D, and SiO. In the \HCOp\ $J=1-0$ line profiles, we detected high
velocity blue- and red-shifted emission in 25 sources ($71\%$), indicating the presence of
protostellar outflows in a large fraction of the targets. Moreover, most of the sources showing 
emission in SiO and SO are associated with high velocity wings in \HCOp, indicating that also in
the low metallicity environment of the OG these molecular species are good
tracers of outflows and protostellar activity. 
We have investigated whether the detection in the various molecules depends on some physical properties
of the sources, such as $R_{\rm GC}$, $d$, $M$, $L$, $L/M$, and $N(\rm H_2)$.
Overall, no clear trends have been found, indicating that the presence of the molecules
analysed in this work do not depend neither on the Galactocentric distance nor on other
physical parameters, even though in some cases the statistics is low and does not allow
us to derive firm conclusions. Similarly, no significant differences have been found between
sources with or without high velocity blue- and red-shifted wings in the \HCOp\ line profiles.
We suggest a tentative difference for SiO, for which the
detected sources are more likely associated with more massive and luminous
objects. Our study clearly supports previous claims that 
organic molecules and tracers of protostellar activity are ubiquitous in the Galaxy. Our results
and the additional, growing evidence that the formation of terrestrial planets is possible also 
at low metallicity, put into question former, stringent definitions of GHZ, in which the capacity of 
the environment to form both organic and other pre-biotic molecules is not taken into account.

\begin{acknowledgements}
F.F. is grateful to the IRAM 30m staff for their precious help during the observations.
This publication has received funding from the European Union Horizon 2020 
research and innovation programme under grant agreement No 730562 (RadioNet).
L.C. and V.M.R. acknowledge support from the Comunidad de Madrid through the 
Atracci\'on de Talento Investigador Modalidad 1 (Doctores con experiencia) Grant 
(COOL:Cosmic Origins of Life; 2019-T1/TIC-15379).
\end{acknowledgements}

{}

\newpage 

\begin{appendix}
\renewcommand{\thefigure}{A-\arabic{figure}}
\renewcommand{\thetable}{A-\arabic{table}}
\setcounter{figure}{0}
\setcounter{table}{0}
\section*{Appendix A: spectrum and line intensities of WB89-437}
\label{appb}

\begin{figure*}
{\includegraphics[width=16cm]{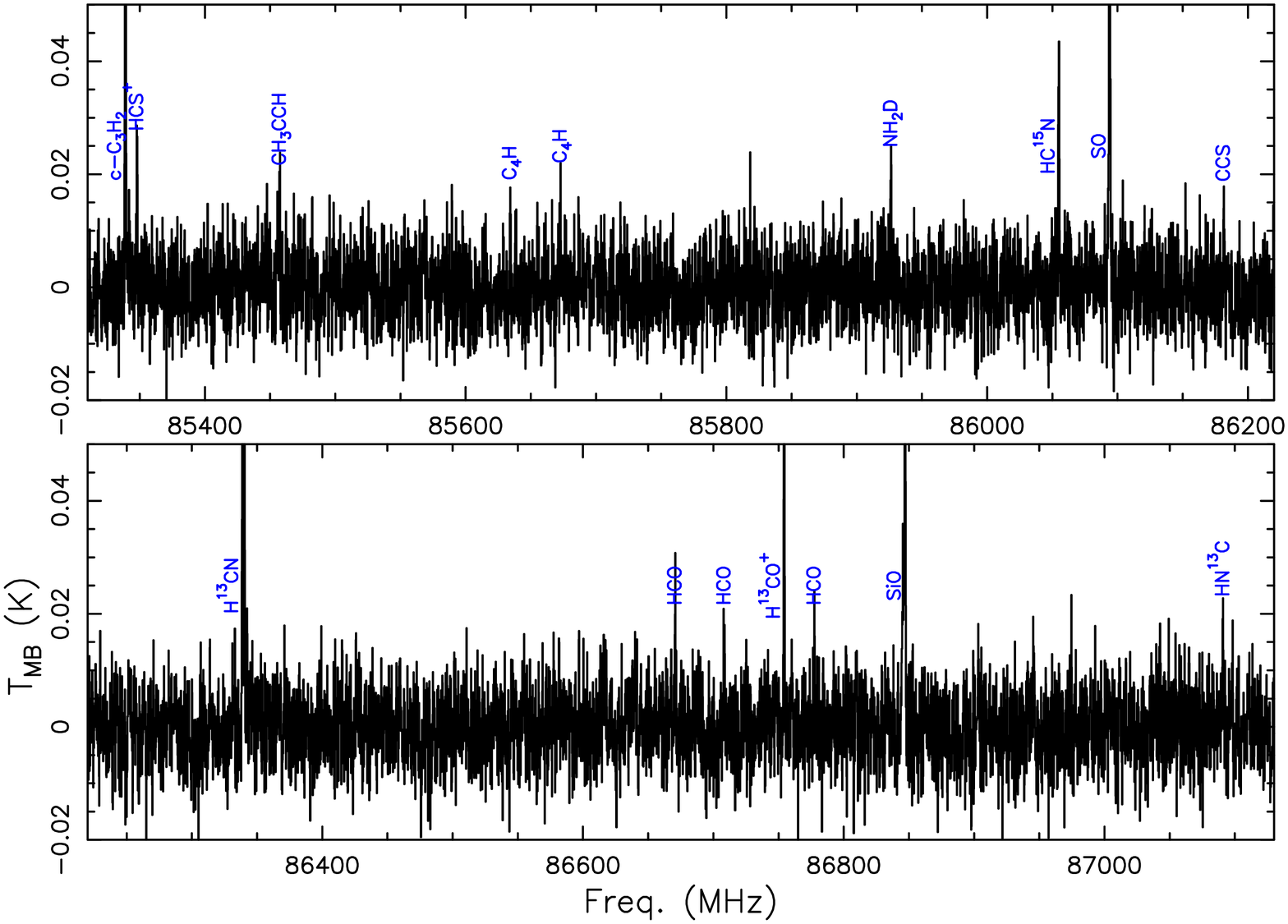}
\includegraphics[width=16cm]{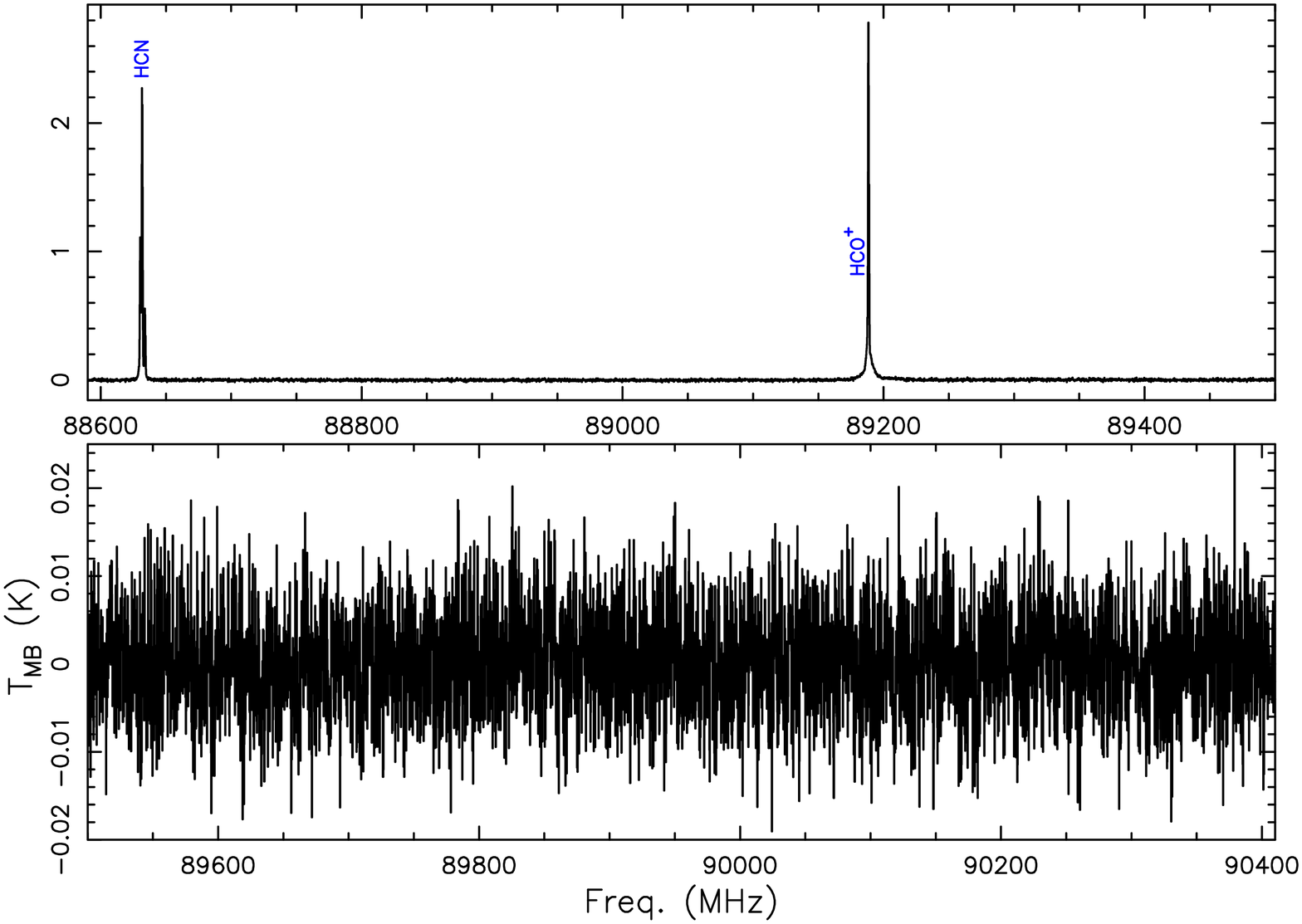}
}
      \caption{Full 3~mm spectrum obtained towards WB89-437. Detected lines of the molecules 
      listed in Table~\ref{tab:detections} are indicated. The spectral parameters of all transitions are given 
      in Table~\ref{tab:spectroscopy}.}
         \label{fig:wb89-437}
\end{figure*}

\begin{figure*}
{\includegraphics[width=16.5cm]{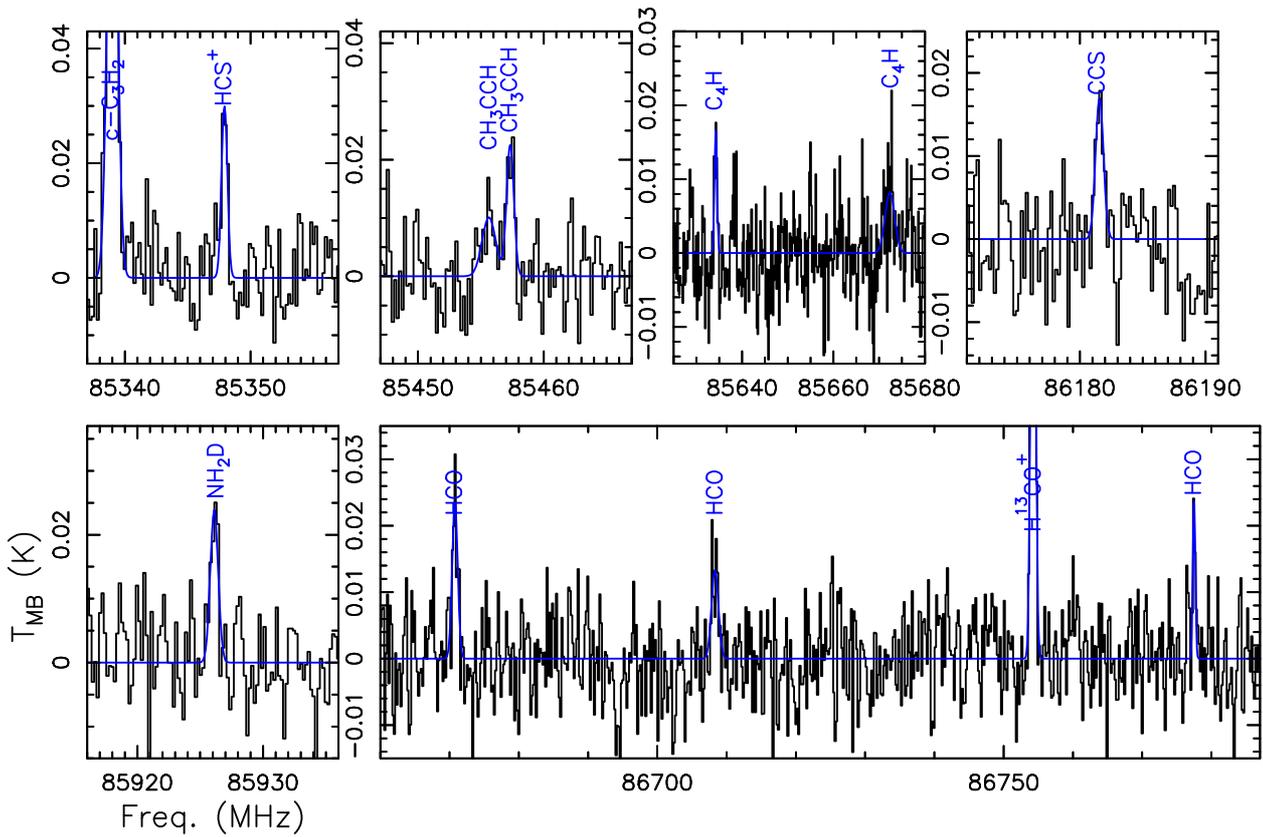}
}
      \caption{Selected spectral windows around faint lines in the spectrum at 3~mm of WB89-437. The
      blue curve in each panel shows the best Gaussian fit to the detected lines.}
         \label{fig:wb89-437-zoom}
\end{figure*}

In this appendix, we show the full spectrum observed at 3~mm towards WB89-437 (Fig.~\ref{fig:wb89-437}),
and zoomed spectral windows around the faintest detected lines (Fig.~\ref{fig:wb89-437-zoom}). 
The intensities of the detected transitions are listed in Table~A-1.

\begin{table*}
\begin{center}
\setlength{\tabcolsep}{1.8pt}
\label{tab:intensities}
\caption{Peak intensities, in K ($T_{\rm MB}$ units) of the transitions detected towards WB89-437 
(Figs.~\ref{fig:wb89-437} and ~\ref{fig:wb89-437-zoom}), obtained from Gaussian fits performed with 
{\sc class}. The calibration errors are indicated in parentheses.}
 \begin{tabular}{cccccc}
\hline \hline
 
 {\it c-}C$_3$H$_2$ & HCS$^+$ & CH$_3$CCH$^{(a)}$  & C$_4$H$^{(b)}$ & CCS  & HCO$^{(c)}$ \\
   \hline
0.1(0.01) & 0.030(0.003) & 0.022(0.02) & 0.017(0.002) & 0.017(0.002) & 0.024(0.003) \\
\hline
   H$^{13}$CO$^+$ & HCN$^{(d)}$ & HCO$^+$ & NH$_2$D & SO & SiO \\
   \hline
0.10(0.01) & 2.3(0.3) & 2.7(0.3) & 0.025(0.003) & 0.17(0.02) & 0.060(0.007) \\
\hline
\hline
\end{tabular}
\end{center}
$^{(a)}$ $K=0$ line (Table~\ref{tab:spectroscopy}); \\
$^{(b)}$ strongest group of unresolved hyperfine components $J=19/2-17/2$ (Table~\ref{tab:spectroscopy}); \\
$^{(c)}$ strongest resolved hyperfine component $J=3/2-1/2$, $F=2-1$ (Table~\ref{tab:spectroscopy}); \\
$^{(d)}$ strongest resolved hyperfine component $F=2-1$ (Table~\ref{tab:spectroscopy}). \\
\end{table*}

\newpage

\renewcommand{\thefigure}{B-\arabic{figure}}
\renewcommand{\thetable}{B-\arabic{table}}
\setcounter{figure}{0}
\setcounter{table}{0}
\section*{Appendix B: Gaussian analysis of the \HCOp\ and {\it c}-C$_3$H$_2$ lines}
\label{appb}
\begin{figure*}
{\includegraphics[width=8.5cm]{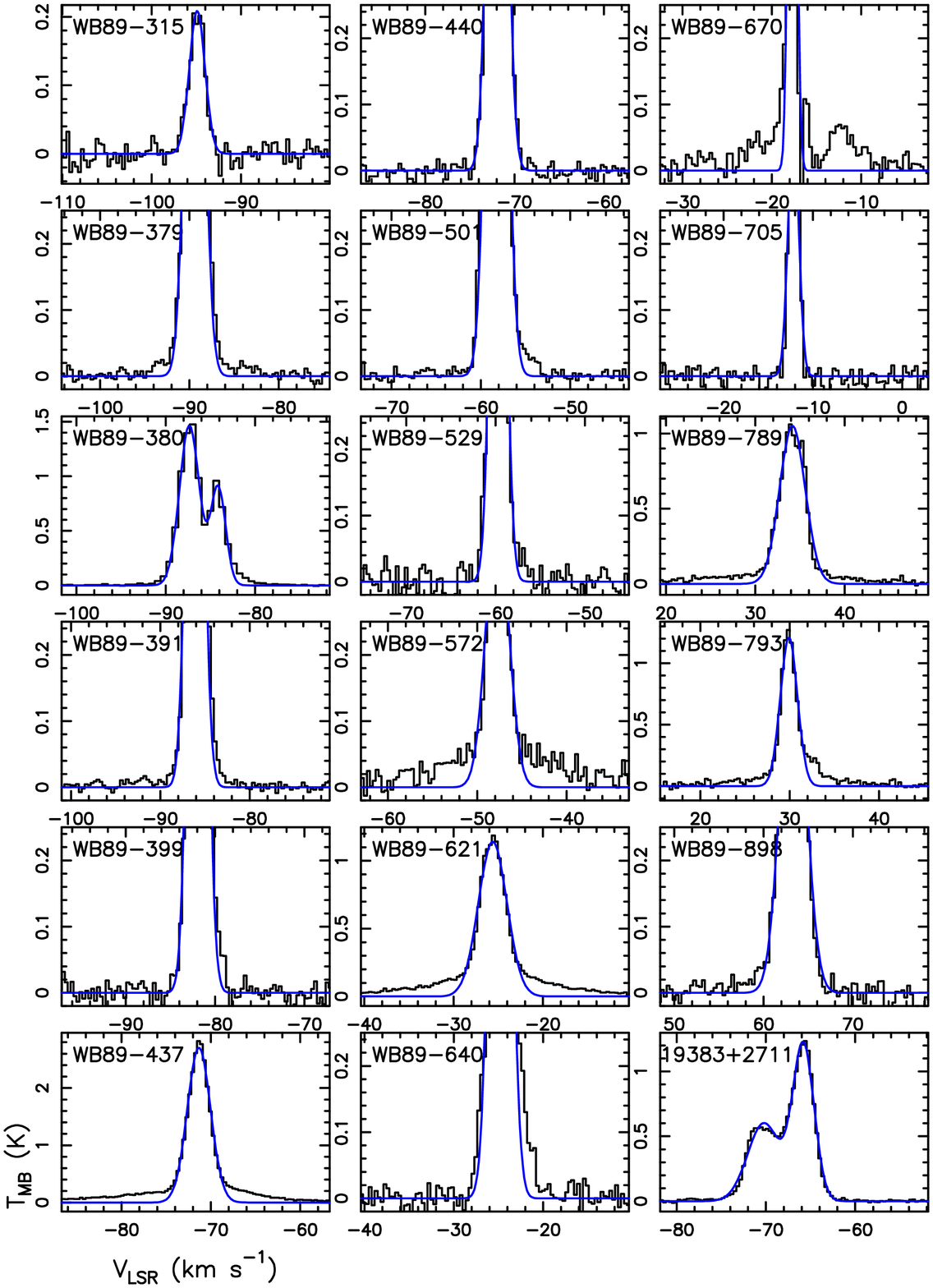}
\includegraphics[width=8.5cm]{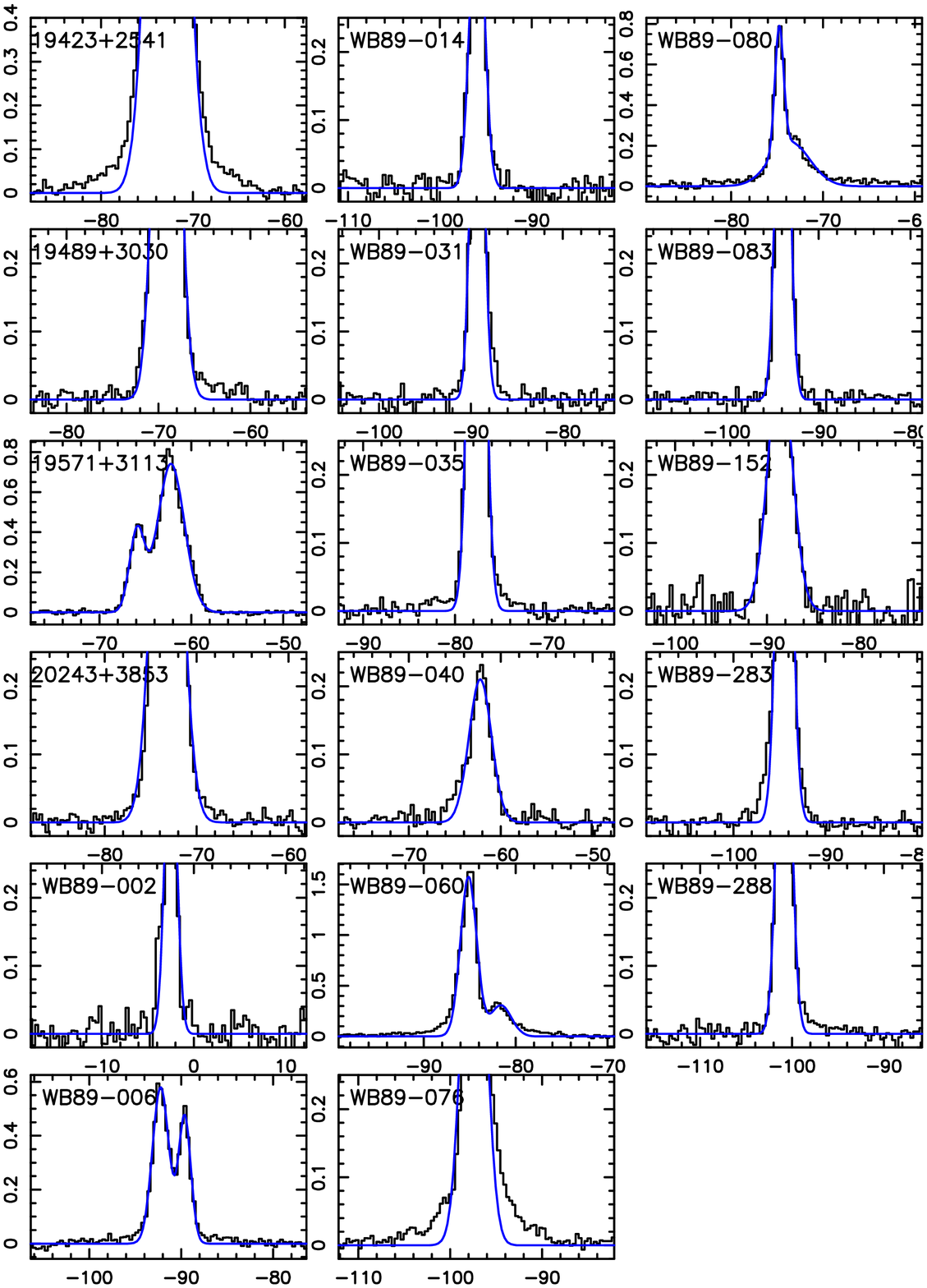}}
      \caption{Spectra of \HCOp\ $J=1-0$ obtained with the IRAM-30m telescope
      towards the sources listed in Table~\ref{tab:sources}. The blue curve in each frame represents the best 
      Gaussian fit to the line. For spectra with two intensity peaks, a double Gaussian fit has been performed.
      In some cases, the y-axis has been reduced to highlight the (tentative) red- and blue-shifted high-velocity
      non-Gaussian wings.   }
         \label{fig:HCOp-Fig1}
\end{figure*}

\begin{figure*}
{\includegraphics[width=8.5cm]{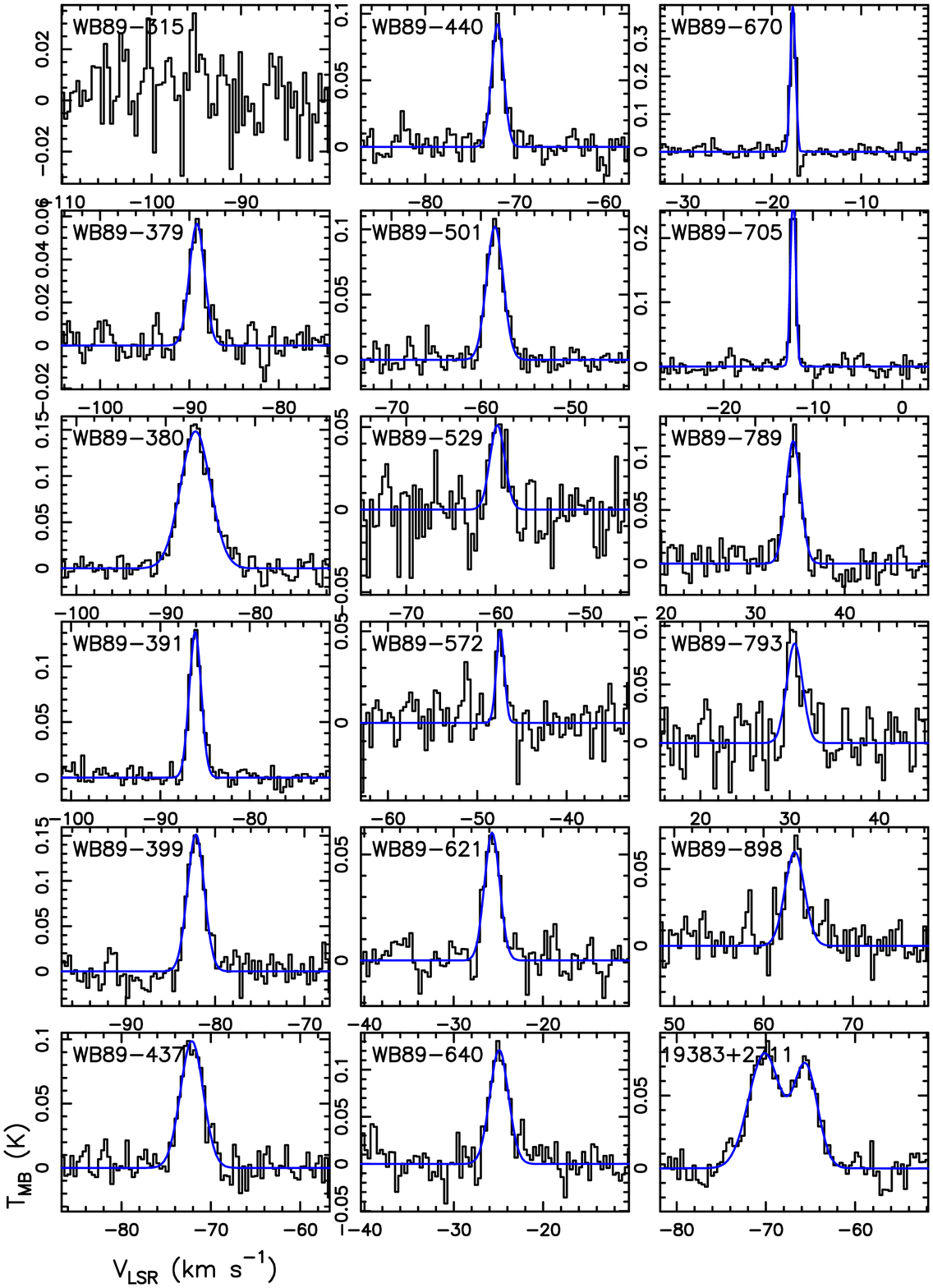}
\includegraphics[width=8.5cm]{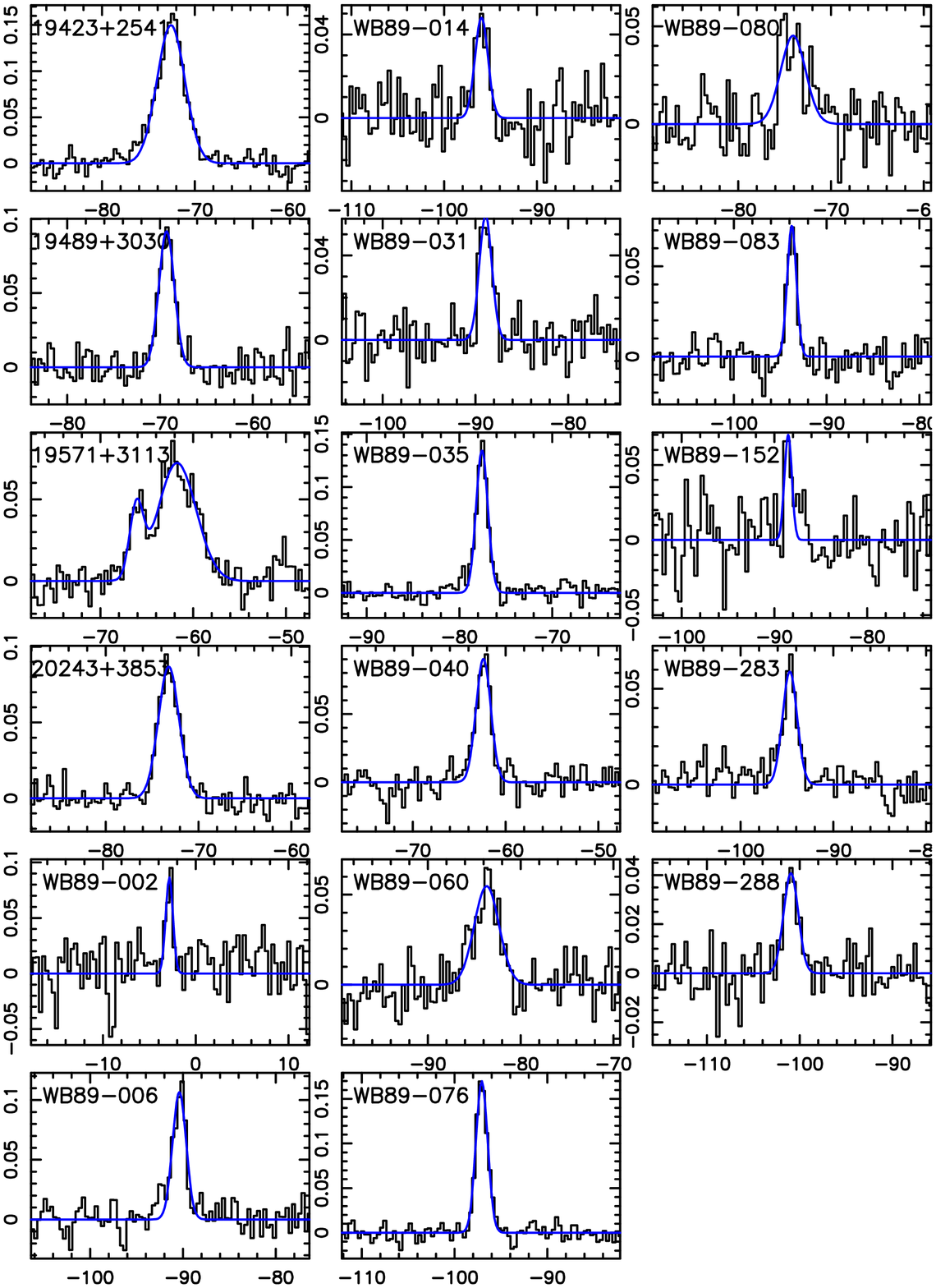}}
      \caption{Same as Fig.~\ref{fig:HCOp-Fig1} for {\it c-}C$_3$H$_2$ $J_{K_a,K_b}=2_{1,2}-1_{0,1}$. 
               }
         \label{fig:C3H2-Fig1}
\end{figure*}

We show in this appendix the spectra of \HCOp\ $J=1-0$ and {\it c}-C$_3$H$_2$ $J_{K_a,K_b}=2_{1,2}-1_{0,1}$ 
(Figs.~\ref{fig:HCOp-Fig1} and \ref{fig:C3H2-Fig1}), analysed in Sects.~\ref{outflows} and \ref{dist}, 
respectively, and the line parameters obtained from Gaussian fits to the lines (Tables~A-1 
and A-2).

\begin{table}
\begin{center}
\setlength{\tabcolsep}{1.8pt}
\label{tab:fits-HCOp}
\caption{Best fit parameters from Gaussian fits to the \HCOp\ $J=1-0$ lines shown in Fig.~\ref{fig:HCOp-Fig1}. 
Uncertainties derived from the fitting procedure are quoted in parentheses.}
 \begin{tabular}{lcccc}
\hline \hline
source       &  $\int T_{\rm MB}{\rm d}V$$^{(1)}$   &  $V_{\rm p}$$^{(2)}$   &  FWHM$^{(3)}$     &  $T_{\rm MB}^{\rm p}$$^{(4)}$   \\
                  &   K km s$^{-1}$                    &   \kms\           &  \kms\       &   K      \\
\hline
WB89--315    &   0.46(0.02)   &  --94.91(0.04)   &   2.07(0.09)   &  0.209    \\
WB89--379    &   2.19(0.01)    & --89.41(0.01)    &  2.20(0.01)    & 0.936     \\
WB89--380    &   4.057(0.001)    & --87.36(0.01)    &  2.60(0.01)    &  1.47     \\
WB89--380-b$^{(a)}$  &     1.868(0.001)  &   --84.14(0.02)  &    2.0(0.2)      &   0.898   \\
WB89--391    &   2.232(0.007)    & --86.12(0.01)    &  1.94(0.01)    &  1.08     \\
WB89--399    &   2.97(0.02)    & --81.94(0.01)    &  2.03(0.01)    &  1.38     \\
WB89--437    &   8.8(0.12)       &  --71.34(0.02)    &  3.06(0.05)    &  2.70     \\
WB89--440    &   2.72(0.01)    & --71.90(0.01)    &  2.07(0.01)    &  1.24     \\
WB89--501    &   2.99(0.01)    & --58.22(0.01)    &  2.28(0.01)    &  1.23     \\
WB89--529    &   1.85(0.02)    & --59.69(0.01)    &  1.82(0.02)    & 0.956     \\
WB89--572    &   1.30(0.03)    & --47.74(0.04)    &  2.9(0.1)    & 0.415     \\
WB89--621    &   4.57(0.08)    & --25.62(0.03)    &  3.76(0.08)    &  1.14     \\
WB89--640    &   3.35(0.02)    & --24.78(0.01)    &  2.27(0.02)    &  1.39     \\
WB89--670    &   1.15(0.02)    & --17.69(0.01)    &  0.97(0.02)    &  1.12     \\
WB89--705    &   0.57(0.01)   &  --12.17(0.02)    &   1.40(0.03)   &  0.381    \\
WB89--789    &   3.77(0.04)    &  34.14(0.02)    &  3.37(0.04)    &  1.05     \\
WB89--793    &   2.90(0.05)    &  29.91(0.02)    &  2.25(0.05)    &  1.21     \\
WB89--898    &   1.95(0.02)    &  63.31(0.02)    &  3.17(0.03)    & 0.576     \\
19423+2541  &     8.48(0.03)  &   --72.79(0.02)  &    3.81(0.02)  &    2.09   \\
19383+2711  &     2.75(0.03)  &   --70.22(0.02)  &    4.29(0.03)  &   0.601   \\
19383+2711-b$^{(a)}$ &     3.60(0.02)  &   --65.81(0.01)  &    2.82(0.01)  &      1.20 \\
19489+3030  &     2.74(0.01)  &   --69.09(0.01)  &    2.70(0.01)  &   0.953   \\
19571+3113  &     0.82(0.01)  &   --65.90(0.01)  &    1.93(0.03)  &    0.400  \\
19571+3113-b$^{(a)}$ &     2.70(0.02   &   --62.21(0.01)  &    3.42(0.03)  &     0.741 \\
20243+3853  &     2.89(0.01)  &   --73.20(0.01)  &    3.33(0.02)  &   0.816   \\
WB89--002    &   0.75(0.02)    &   --2.50(0.02)   &   1.48(0.05)   &  0.480    \\
WB89--006    &   1.30(0.02)    &  --92.23(0.02)    &  2.10(0.04)   & 0.579     \\
WB89--006-b$^{(a)}$  &     0.78(0.02)  &    --89.61(0.02) &   1.55(0.05)   &    0.470  \\
WB89--014    &   0.88(0.02)    &  --95.93(0.02)   &   1.79(0.04)   &  0.464    \\
WB89--031    &   1.12(0.01)    & --89.29(0.01)    &   1.53(0.02)    & 0.690     \\
WB89--035    &   1.84(0.01)    & --77.62(0.01)    &   1.86(0.01)    & 0.930     \\
WB89--040    &   0.64(0.02)   &  --62.22(0.03)   &    2.88(0.09)   &  0.210    \\
WB89--060    &   3.47(0.04)    & --85.13(0.01)    &   2.07(0.03)    &  1.57     \\
WB89--060-b$^{(a)}$  &     0.88(0.04)  &  --81.61(0.06) &     2.7(0.2)     &    0.307  \\
WB89--076    &   2.00(0.03)    & --97.43(0.02)    &   2.74(0.04)    & 0.685     \\
WB89--080    &   0.71(0.03)   &  --74.77(0.01)   &    1.10(0.03)   &  0.603    \\
WB89--080-b$^{(a)}$  &     1.18(0.04)  &   --73.61(0.08)  &    5.1(0.2)      &   0.218   \\
WB89--083    &   1.210(0.008)    & --93.91(0.04)   &  1.59(0.01)    & 0.715     \\
WB89--152    &   1.21(0.03)    & --88.56(0.04)    &  3.1(0.1)    & 0.368     \\
WB89--283    &   1.51(0.01)    & --94.45(0.01)    &  1.78(0.01)    & 0.796     \\
WB89--288    &   1.15(0.01)    & --100.9(0.01)    &  1.74(0.02)    & 0.620     \\
\hline
\end{tabular}
\end{center}
$^{(1)}$ line integrated intensity; \\
$^{(2)}$ peak velocity; \\
$^{(3)}$ full width at half maximum; \\
$^{(4)}$ intensity peak; \\
$^{(a)}$ second velocity feature. \\
\end{table}

\begin{table}
\begin{center}
\setlength{\tabcolsep}{1.8pt}
\label{tab:fits-C3H2}
\caption{Best fit parameters from Gaussian fits to the {\it c-}C$_3$H$_2$ $J_{K_a,K_b}=2_{1,2}-1_{0,1}$ lines shown 
in Fig.~\ref{fig:C3H2-Fig1}. Uncertainties derived from the fitting procedure
are quoted in parentheses.}
 \begin{tabular}{lcccc}
 \hline \hline
 source       &  $\int T_{\rm MB}{\rm d}V$$^{(1)}$  &  $V_{\rm p}$$^{(2)}$  &  FWHM$^{(3)}$    &  $T_{\rm MB}^{\rm p}$$^{(4)}$  \\
                  &   K km s$^{-1}$                    &   \kms\           &  \kms\       &   K      \\
\hline
 WB89--315    &  $\leq 0.04$  &  --      & --     &  $\leq 0.03$   \\
 WB89--379    &   0.121(0.007) &  --89.16(0.06)  &  2.0(0.2)      &  0.057   \\
 WB89--380    &   0.63(0.02) &  --86.68(0.04)  &  4.0(0.1)      &  0.148       \\
 WB89--391    &   0.220(0.006) &  --86.10(0.02)  &  1.58(0.05)  &  0.131       \\
 WB89--399    &  0.36(0.02)  &  --82.15(0.05)  &  2.2(0.1)      &  0.151       \\
 WB89--437    &  0.32(0.01)  &  --72.14(0.06)  &  3.0(0.2)      &  0.099   \\
 WB89--440    &  0.16(0.01)  &  --71.88(0.05)  &  1.6(0.1)      &  0.092   \\
 WB89--501    &   0.24(0.01) &  --58.43(0.04)  &  2.2(0.1)      &  0.102       \\
 WB89--529    &  0.11(0.02)  &  --59.8(0.2)      &  2.0(0.4)      &  0.051   \\
 WB89--572    &  0.06(0.01) &  --47.4(0.1)      &  1.0(0.3)      &  0.050   \\
 WB89--621    &   0.13(0.01) &  --25.68(0.07)  &  2.0(0.2)      &  0.060   \\
 WB89--640    &  0.32(0.02)  &  --24.93(0.08)  &  2.5(0.2)      &  0.121       \\
 WB89--670    &  0.28(0.01)  &  --17.65(0.01)  & 0.68(0.02)  &  0.387       \\
 WB89--705    &  0.196(0.008)  &  --12.20(0.01)  & 0.67(0.03)  &  0.275       \\
 WB89--789    &  0.23(0.01)  &   34.25(0.05)  &  1.9(0.1)      &  0.114       \\
 WB89--793    &  0.19(0.03)  &   30.5(0.2)      &  2.1(0.6)      &  0.085   \\
 WB89--898    &  0.17(0.02)  &   63.5(0.1)      &  2.6(0.4)      &  0.061   \\
 19423+2541  &  0.56(0.01)  &  --72.58(0.04)  &  3.5(0.1)  &  0.150       \\
 19383+2711  &  0.35(0.03)  &  --70.2(0.2)      &  4.2(0.4)      &  0.079   \\
 19383+2711-b$^{(a)}$  &  0.23(0.03)   &  --65.6(0.2)  & 3.1(0.3)  &  0.069 \\
 19489+3030  &  0.19(0.01)  &  --69.29(0.05)  &  2.0(0.1)      &  0.092   \\
 19571+3113  &  0.37(0.02)  &  --61.7(0.1)     &  4.8(0.3)      &  0.072   \\
 19571+3113-b$^{(a)}$  &  0.08(0.01)   &  --66.2(0.1)  & 1.8(0.3)   &  0.043 \\
 20243+3853  &  0.24(0.01)  &  --73.21(0.05)  &  2.6(0.1)      &  0.087   \\
 WB89--002    &  0.08(0.02)  &  --2.8(0.1)       & 0.9(0.2)      &  0.087   \\
 WB89--006    &  0.20(0.01)  &  --90.38(0.05)  &  1.7(0.2)      &  0.107       \\
 WB89--014    &  0.08(0.01)  &  --96.0(0.1)    &  1.6(0.3)      &  0.048   \\
 WB89--031    &  1.00(0.01)  &  --88.89(0.08)  &  1.6(0.2)      &  0.059   \\
 WB89--035    &  0.21(0.08)  &  --77.56(0.03)  &  1.50(0.07)  &  0.135       \\
 WB89--040    &  0.18(0.01)  &  --62.38(0.05)  &  1.9(0.1)      &  0.090   \\
 WB89--060    &  0.18(0.02)  &  --83.7(0.2)      &  3.1(0.3)      &  0.055   \\
 WB89--076    &  0.270(0.008)  &  --97.07(0.02)  &  1.49(0.05)  &  0.170       \\
 WB89--080    &  0.15(0.02)  &  --74.1(0.2)      &  3.2(0.5)      &  0.045   \\
 WB89--083    &  0.090(0.007) &   --93.76(0.04) &   1.2(0.1) &   0.073  \\
 WB89--152    &  0.07(0.02)  &  --88.5(0.2)      & 1.0(0.5)      &  0.070   \\
 WB89--283    &   0.12(0.01) &   --94.69(0.06) &   1.9(0.2)     &   0.059  \\
 WB89--288    &   0.08(0.01) &   --101.0(0.1)     &   1.8(0.3)     &   0.041  \\
\hline
\end{tabular}
\end{center}
$^{(1)}$ line integrated intensity; \\
$^{(2)}$ peak velocity; \\
$^{(3)}$ full width at half maximum; \\
$^{(4)}$ intensity peak; \\
$^{(a)}$ second velocity feature. \\
\end{table}

\newpage 

\renewcommand{\thefigure}{C-\arabic{figure}}
\renewcommand{\thetable}{C-\arabic{table}}
\setcounter{figure}{0}
\setcounter{table}{0}
\section*{Appendix C: kinematic distances derived from the rotation curve of \citet{reid19}.}
\label{appc}

\begin{table}
\begin{center}
\label{tab:distances-reid}
\caption{Kinematic Galactocentric and heliocentric distances derived from the rotation curve of \citet{reid19}.}
 \begin{tabular}{lcc}
 \hline \hline
 source   & $R^{new}_{\rm GC}$$^{(1)}$ & $d$$^{(1)}$ \\
              &   kpc               & kpc \\
\hline
    WB89-315  &     14.63  &  8.72   \\
    WB89-379  &    14.66 &   8.21   \\
    WB89-380  &    14.35  &  7.86  \\
    WB89-391  &    14.43   & 7.86   \\
    WB89-399  &    14.34   & 7.55   \\
    WB89-437  &    14.02  &  6.79  \\
    WB89-440  &    14.04  &  6.80   \\
    WB89-501  &    13.92  &  6.22  \\
    WB89-529  &    15.58 &   7.79   \\
    WB89-572  &    15.74 &   7.71   \\
    WB89-621  &    15.44  &  7.17   \\
    WB89-640  &     14.1   & 5.84   \\
    WB89-670  &    17.04   & 8.72    \\
    WB89-705  &    14.63   & 6.30    \\
    WB89-789  &    18.66 &   10.49  \\
    WB89-793  &    16.57 &   8.38  \\
    WB89-898  &    14.93 &   7.42  \\
  19423+2541  &    12.13  &  13.60   \\
  19383+2711  &    12.28  &  13.63  \\
  19489+3030  &    11.84  &  12.34  \\
  19571+3113  &    11.26 &   11.27  \\
  20243+3853  &    11.84  &  10.37  \\
    WB89-002  &    8.369 &   1.54  \\
    WB89-006  &    13.02   & 10.55   \\
    WB89-014  &    13.49 &   10.74  \\
    WB89-031  &    12.91  &  10.13  \\
    WB89-035  &    12.03  &  8.67   \\
    WB89-040  &    11.04  &  7.12  \\
    WB89-060  &     12.5 &   8.59   \\
    WB89-076  &    13.64  &  10.05  \\
    WB89-080  &    11.81 &   7.60  \\
    WB89-083  &    13.35  &  9.57   \\
    WB89-152  &    13.07  &  8.24  \\
    WB89-283  &    14.24  &  8.60   \\
    WB89-288  &    14.98  &  9.47    \\
 \hline
\end{tabular}
\end{center}
$^{(1)}$ We adopted the same Longitude and $V_{\rm p}$ (for sources having two velocity features, only the main one is
used) as in Table~\ref{tab:newdist}. \\
\end{table}

\end{appendix}

\end{document}